\documentclass[superscriptaddress,prb,twocolumn,floatfix,aps,longbibliography]{revtex4-2}
\usepackage{graphicx}
\usepackage{amsmath,amssymb,physics,dsfont}
\usepackage{mathrsfs}
\usepackage{color}
\usepackage{hyperref}
\usepackage{microtype}
\usepackage{braket}
\usepackage{defs-private}
\usepackage{tabularray}
\begin{document}

\title{Adaptive variational quantum dynamics simulations with compressed circuits and fewer measurements}

\author{Feng Zhang}
\email{fzhang@ameslab.gov}
\affiliation{Ames National Laboratory, U.S. Department of Energy, Ames, Iowa 50011, USA}
\affiliation{Department of Physics and Astronomy, Iowa State University, Ames, Iowa 50011, USA}

\author{Cai-Zhuang Wang}
\affiliation{Ames National Laboratory, U.S. Department of Energy, Ames, Iowa 50011, USA}
\affiliation{Department of Physics and Astronomy, Iowa State University, Ames, Iowa 50011, USA}

\author{Thomas Iadecola}
\affiliation{Ames National Laboratory, U.S. Department of Energy, Ames, Iowa 50011, USA}
\affiliation{Department of Physics and Astronomy, Iowa State University, Ames, Iowa 50011, USA}

\author{Peter P. Orth}
\affiliation{Ames National Laboratory, U.S. Department of Energy, Ames, Iowa 50011, USA}
\affiliation{Department of Physics and Astronomy, Iowa State University, Ames, Iowa 50011, USA}
\affiliation{Department of Physics, Saarland University, 66123 Saarbr\"ucken, Germany}

\author{Yong-Xin Yao}
\email{ykent@iastate.edu}
\affiliation{Ames National Laboratory, U.S. Department of Energy, Ames, Iowa 50011, USA}
\affiliation{Department of Physics and Astronomy, Iowa State University, Ames, Iowa 50011, USA}

\begin{abstract}
The adaptive variational quantum dynamics simulation (AVQDS) method performs real-time evolution of quantum states using automatically generated parameterized quantum circuits that often contain substantially fewer gates than Trotter circuits. 
Here we report an improved version of the method, which we call AVQDS(T), by porting the Tiling Efficient Trial Circuits with Rotations Implemented Simultaneously (TETRIS) technique. 
The algorithm adaptively adds layers of disjoint unitary gates to the ansatz circuit so as to keep the McLachlan distance, a measure of the accuracy of the variational dynamics, below a fixed threshold.
We perform benchmark noiseless AVQDS(T) simulations of quench dynamics in local spin models, and compare with an alternative adaptive variational approach on quantum resource requirement. Quantum dynamical simulations implementing realistic noise channels are also reported.
Finally, we propose a way to substantially alleviate the measurement overhead of AVQDS(T) while maintaining high accuracy by synergistically integrating quantum circuit calculations on quantum processing units with classical calculations using, e.g., tensor networks to evaluate the quantum geometric tensor.
We showcase that this approach enables AVQDS(T) to deliver more accurate results than simulations using a fixed ansatz of comparable final depth for a significant time duration with fewer quantum resources.
\end{abstract}

\maketitle

\section{Introduction}
Quantum many-body systems have been a fruitful testbed for quantum computing~\cite{feynman82qc, lloyd1996, Abrams97simulation, pea_lloyd, somma03quantum, asp_ipea, kassal08polynomial, rmp_qs, TroyerQCMB, Trotter_dynamics_Lawrence, Trotter_dynamics_Knolle,Chen2022, rmp_qcc, miessen2023quantum}, since they admit a natural mapping onto a qubit-based computing architecture~\cite{feynman82qc}. 
In the near term, practical quantum algorithms need to be tailored for noisy intermediate-scale quantum (NISQ) hardware~\cite{nisq} with limited qubits and coherence times. 
A typical example of such resource-efficient algorithms is the variational quantum eigensolver (VQE), which was introduced to solve static quantum problems such as finding the ground and excited states of a Hamiltonian by employing the variational principle~\cite{peruzzoVariationalEigenvalueSolver2014, vqe_theory, vqe_pea_h2, hardware_efficient_vqe, qcc_scott2018, FengVQE}. 
In VQE, parameterized trial states are prepared and measured on quantum processing units (QPUs), and a target cost function is calculated based on the measurement outcomes. 
The quantum circuits are then instructed by an optimization algorithm that runs on a classical computer on how to reach the optimal quantum state by updating the variational parameters. 
The variational ansatz plays a crucial role in bounding the accuracy of VQE. 
The problem-agnostic hardware efficient ansatz can in principle boost its expressibility by increasing the number of circuit layers, but practical applications are complicated by the presence of barren plateaus~\cite{mcclean2018barren,Larocca2024}, where the cost function gradient vanishes exponentially with increasing number of qubits. 
The unitary coupled cluster ansatz with single and double excitations (UCCSD) is reasonably accurate to describe equilibrium ground states in quantum chemistry~\cite{vqe_uccsd}, but the associated parameterized circuits can still be rather deep even for small molecules.
One way of compressing the circuits while improving their ability to represent the ground state is to construct problem-specific ans\"atze by adaptively adding parameterized unitaries according to figures of merit such as the cost function gradient~\cite{grimsleyAdaptiveVariationalAlgorithm2019} and state overlap~\cite{feniou2023overlapav}. 
Examples include the adaptive derivative-assembled pseudo-trotter ansatz (ADAPT) approach and its variant known as qubit-ADAPT-VQE~\cite{grimsleyAdaptiveVariationalAlgorithm2019, MayhallQubitAVQE, feniou2023overlapav}.

Variational approaches with fixed ans\"atze have also been used to study the dynamics of quantum systems(VQDS)~\cite{theory_vqs, Endo20variational, cirstoiu2020variational, benedetti2020hardware, chen19demonstration, Berthusen:2022}. 
Here, it becomes more challenging to construct the ans\"atze for dynamics simulations, because they are expected to faithfully represent the dynamical quantum state during the entire time evolution. 
For this reason, the adaptive variational quantum dynamics simulation (AVQDS) approach has been proposed~\cite{AVQDS}, which dynamically expands the variational ansatz to keep a figure of merit for the variational dynamics, known as the McLachlan distance $\mathcal L^2$, a measure of the fidelity of the time-evolved variational state, below a fixed threshold along the time-evolution path~\cite{mclachlan64variational, theory_vqs}. 
Compared with standard first-order Trotterized circuits for dynamics simulations of quantum spin models, it has been demonstrated that AVQDS can significantly reduce the circuit complexity as measured by the number of two-qubit entangling gates~\cite{AVQDS, mootz2023twodimensional}. Alternative algorithms for compressing quantum circuits in dynamics simulations include projected variational quantum dynamics (pVQD)~\cite{Barison2021efficientquantum} and its adaptive variant~\cite{Linteau2024adaptiveprojectedVQD}, as well as variational fast forwarding (VFF) and the subspace approach~\cite{cirstoiu2020variational, gibbs2021longtime, Heya2023subspaceVQS, cerezo2021variational}. The adaptive pVQD algorithm iteratively constructs and optimizes a variational ansatz to maintain a high overlap with the quantum state evolved using one or more Trotter steps along the dynamical path. While both AVQDS and pVQD are suited for dynamics simulations with either fixed or time-dependent Hamiltonians, VFF and the subspace approach are tailored for sudden quench dynamics under time-independent Hamiltonians. The common underlying idea is to design variational circuits that approximately diagonalize the Hamiltonian evolution operator \(e^{-i\hat{H}t}\) either across the entire Hilbert space or within a low-energy subspace, thereby simplifying the simulation.

In the original ADAPT-VQE and AVQDS methods, the parameterized unitary gates are added iteratively and one at a time when the ansatz is expanded. 
In general, each unitary only acts on a fraction of the qubits, leaving many of them idling. 
Adding the next unitary without taking the idle qubits into consideration can cause unnecessary deepening of the quantum circuit, making it suboptimal to run on QPUs. 
As a further improvement, Anastasiou {\it et al.} introduced the tiling efficient trial circuits with rotations implemented simultaneously (TETRIS) technique~\cite{Anastasiou2022} to the original ADAPT-VQE method as an improved prescription for adding unitaries to the ansatz~\cite{MayhallQubitAVQE}. TETRIS begins like the standard ADAPT protocol by appending to the ansatz a unitary producing the largest energy gradient. The set of possible unitaries is characterized by a predefined operator pool consisting of the allowed generators of a unitary rotation.
However, instead of proceeding with the VQE, the TETRIS approach keeps checking other unitaries from the pool with energy gradients ranked in descending order and adds the ones that act on disjoint sets of idling qubits until a full gate layer is obtained. 
In addition to the expected advantage of generating shallower ansatz circuits, the TETRIS technique was found to reduce the number of ansatz reoptimizations entailed in ADAPT-VQE calculations of small molecules, thereby speeding up convergence~\cite{Anastasiou2022}.

In this work we harness the TETRIS technique for quantum dynamics simulations by incorporating it into the adaptive ansatz expansion subroutine of AVQDS, resulting in a modified version of the algorithm that we denote AVQDS(T). We apply AVQDS(T) to quench dynamics simulations of local spin models and show that this procedure greatly reduces the quantum circuit depth with negligible overhead.
Meanwhile, we present the implementation of an approach based on eigenvalue truncation~\cite{hansen1990trucated,gacon2023ieee} to solve the linear equations of motion for propagating the variational parameters, and demonstrate that this method can be advantageous over standard techniques like Tikhonov regularization in the presence of noise.

Furthermore, we propose an approach to synergistically integrate classical and quantum resources to address one of the main challenges of variational approaches like AVQDS.
Specifically, AVQDS and related approaches all rely on measurements of the quantum geometric tensor (QGT) to propagate the variational parameters, necessitating the evaluation of a number of circuits that grows quadratically with the number of variational parameters~\cite{theory_vqs}.
The pivotal idea is to capitalize on the observation that the quantum resource-hungry QGT can always be partitioned into an inner block, which can be evaluated efficiently and accurately using classical approaches like tensor networks (TNs), and the remaining components which are beyond the reach of TNs and therefore must be measured on QPUs.
Finally, we showcase that with this implementation AVQDS(T) can deliver more accurate results than VQDS simulations using a comparable fixed ansatz over a significant time duration, while maintaining lower computational costs. 
The idea of hybrid QPU-TN evaluations of the QGT can be naturally extended to ground state preparation using imaginary time evolution~\cite{VQITE, AVQITE} and simulations of mixed states and open quantum systems~\cite{theory_vqs, Chen2024_avqite_open}.
This allows the variational algorithms to maximally leverage the computational power of TNs and only resort to QPUs for the classically challenging part of the calculation, making them promising to achieve quantum utility in large-scale applications.

The paper is organized as follows. We start by describing the AVQDS(T) method in Sec.~\ref{sec: method}. 
In Sec.~\ref{sec: model}, we describe the models used in our benchmark calculations---namely, the one-dimensional transverse-field Ising model, mixed-field Ising model, and Heisenberg model---and review the standard Trotter decomposition approach and the Hamiltonian variational ansatz for VQDS. 
In Sec.~\ref{sec: results}, we start with a comparison of different methods for solving the linear equations of motion on both noiseless and noisy simulators, followed by a detailed benchmark of AVQDS(T) on the models described in Sec.~\ref{sec: model}. 
In Sec.~\ref{sec: qpu-tn}, we discuss how to leverage classical computations to reduce the measurement overhead in AVQDS(T) calculations. 
We conclude our work with an outlook in Sec.~\ref{sec: conclusion}.

\section{Method}
\label{sec: method}
\subsection{Variational Quantum Dynamics Simulations}
For completeness, we begin with a summary of the VQDS algorithm~\cite{theory_vqs}.
The exact time evolution of the density matrix $\rho$ of a quantum state is governed by the von-Neumann equation
\be
\frac{d\rho}{d t} = L[\rho],
\ee
where $L[\rho] = -i[\h, \rho]$ for closed systems described by a Hamiltonian $\h$.
Below we specialize to the case where $\rho=\ket{\Psi}\bra{\Psi}$ is a pure state.
In the VQDS approach, the density matrix is parameterized by a time-dependent vector $\bth(t)$ of $N_{\bth}$ real parameters $\theta_1(t),\theta_2(t),\cdots,\theta_{N_{\bth}}(t)$: $\rho[\bth(t)] = \ket{\Psi[\bth(t)]} \bra{\Psi[\bth(t)]}$. The evolution of the parameterized state is determined by $\bth(t)$ through the equations of motion: 
\be
\sum_\nu M_{\mu \nu}\dot{\theta}_\nu = V_\mu, \label{eq: eom}
\ee
which are derived by minimizing the McLachlan distance $\Lag^2$ between the exact and the variational time evolution~\cite{mclachlan64variational}:
\bea
\Lag^2 &\equiv&\norm{\sum_\mu \frac{\partial \rho[\bth]}{\partial \theta_\mu} \dot{\theta}_\mu - L[\rho]}^2_F,
\eea
where $\norm{\cdot}_F$ denotes the Frobenius norm. 
The symmetric matrix $M$ in Eq.~\eqref{eq: eom} is defined as the real part of the quantum geometric tensor:
\be
M_{\mu \nu} = \Re\left[\frac{\partial \bra{\Psi}}{\partial \theta_\mu} (1-\ket{\Psi}\bra{\Psi})\frac{\partial \ket{\Psi}}{\partial \theta_\nu}\right], \label{eq: M}
\ee
and the vector $V$ is given by
\be
V_\mu = \Im\left[\frac{\partial \bra{\Psi}}{\partial \theta_\mu} (1-\ket{\Psi}\bra{\Psi})\h\ket{\Psi}\right]. \label{eq: V}
\ee
\begin{figure*}[t]
	\centering
	\includegraphics[width=\textwidth]{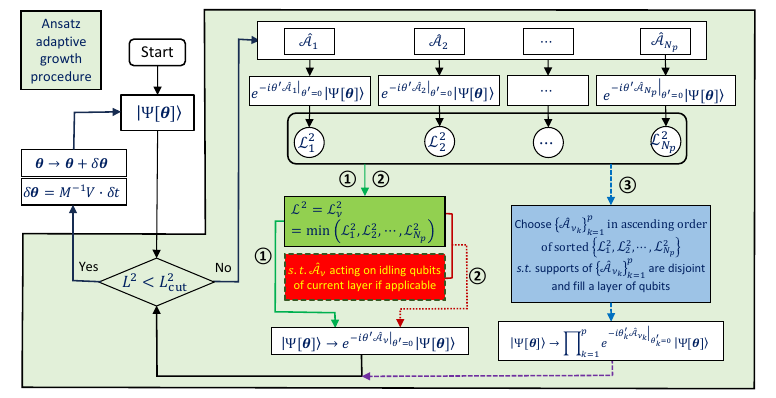}
        \vspace{0cm}
	\caption{
	\textbf{Flowchart illustrating three methods for the adaptive growth procedure of the ansatz in AVQDS.} In Method 1, the single unitary that maximally reduces the McLachlan distance $\mathcal L^2$ is chosen. In Method 2, an additional constraint is enforced such that the unitary must act exclusively on the idling qubits of the current circuit layer if any are available. While only one unitary is selected in Methods 1 and 2, Method 3 allows multiple unitaries to be chosen provided that they act on disjoint sets of qubits to fill out a full circuit gate layer.
	}
	\label{adgp}
\end{figure*}
The explicit dependence of $\ket{\Psi}$ on $\bth(t)$ has been omitted for simplicity. 

In Appendix~\ref{appendix:solvers}, we discuss multiple approaches for solving the equations of motion given in Eq.~\ref{eq: eom}. In the results presented below, we use the truncation method with $\varepsilon=10^{-6}$ and  $\varepsilon=10^{-3}$ for noiseless and noisy simulations, respectively.
\subsection{Adaptive Variational Quantum Dynamics Simulations (AVQDS)}
We consider variational ans{\"a}tze of the general form 
\be
\ket{\Psi[\bth]} = \prod_{\mu=1}^{N_{\bth}} e^{-i\theta_\mu \hat{\A}_\mu} \ket{\Psi_0}, \label{eq: ansatz}
\ee
where $\theta_\mu$ are variational parameters, $\hat{\A}_\mu$ are Hermitian operators selected from a pre-determined pool $\mathscr{P}=\{\hat{\A}_i\}_{i=1}^{N_p}$ made of $N_p$ Pauli strings, and $\ket{\Psi_0}$ is a reference state. 
Specifically, we adopt a pool comprising all the non-identity Pauli strings present in the Hamiltonian of the studied model.
The unitaries $e^{-i\theta_\mu \hat{\A}_\mu}$ can be implemented on quantum devices with a combination of one-qubit rotation gates and two-qubit entangling gates~\cite{nielsen2002quantum}. 
In AVQDS, the ansatz is iteratively expanded during the evolution by adding new operators whenever the McLachlan distance exceeds a threshold value $\mathcal L^2_{\text{cut}}$~\cite{AVQDS}. 
At each time step, one first solves Eq.~\eqref{eq: eom} to obtain $\dot{\bth}$ and the corresponding McLachlan distance $\mathcal L^2=2\left(\text{var}[\h]-V^\dagger\dot{\bth}\right)$, where $\text{var}[\h]=\Av{\Psi(\bth)}{\hat{H}^2}-\left[\Av{\Psi(\bth)}{\h}\right]^2$ is the energy variance of the variational state. 
If $\mathcal L^2<\mathcal L^2_{\text{cut}}$, indicating that the current ansatz still performs well, one proceeds to update the variational state by incrementing the variational parameters via the Euler method: $\Delta\bth=\dot{\bth}\Delta t$~\cite{iserles2009ode}. 
Otherwise, an adaptive ansatz growth procedure is triggered to improve the ansatz expressivity, as illustrated in Fig.~\ref{adgp}. 
The procedure starts by calculating a ``score" for each generator $\hat{\A}_i$ in the pool, defined as the reduction in the McLachlan distance by appending the associated unitary to the current ansatz. 
Specifically, to evaluate the score of a unitary $e^{-i\theta\hat{\A}}$, it is first appended to the ansatz with $\theta$ set to zero. 
This leaves the state unchanged, but increases the dimension of the parameter space $N_{\bth}$ by 1. 
Accordingly, $M$ is augmented with an extra row and column, and an extra element is appended to $V$ (note that the gradient with respect to $\theta$ generically does not vanish even if $\theta$ is set to zero). 
Because $M$ is symmetric, only the new column of $M$ and the new element of $V$ need to be evaluated.
The McLachlan distance $\mathcal L^2$ for the updated ansatz can then be obtained, which is mathematically guaranteed to decrease or stay the same. 
With the score of each unitary evaluated, a set of unitaries is chosen (in a manner described below) to be appended to the ansatz so as to maximally reduce $\mathcal L^2$. 
This process is iterated until a satisfactory $\mathcal L^2 < \mathcal L^2_\textrm{cut}$ is obtained.

Three methods can be adopted for choosing new unitaries to expand the ansatz. 
Method 1, used in the original AVQDS~\cite{AVQDS}, chooses the unitary with the best score at each iteration, leaving most qubits idle. 
In contrast, we propose two new methods, 2 and 3, that take the circuit depth explicitly into account and represent two implementations of the TETRIS approach within AVQDS.
In Method 2, a unitary with the best score is selected subject to the requirement that the unitary must act only on the idling quits of the current circuit layer. 
A new layer is started only if there are no idling qubits in the current layer or if all the unitaries acting on idling qubits fail to deliver any appreciable reduction in $\mathcal L^2$. 
This practice guarantees that the circuit grows layer by layer in a compact way. 
Because of the additional constraint, in principle Method 2 could require more iterations than Method 1 to achieve $\mathcal L^2 < \mathcal L_\textrm{cut}^2$. 
Method 3 avoids this potential increase in the number of iterations by adding $p\ge1$ unitaries at each iteration. To do so, it ranks all unitaries according to their score and proceeds down the list, adding each unitary provided that it does not act on qubits that have already been acted upon by the previous unitaries.
Thus, Method 3 introduces a new full circuit layer at each iteration, in contrast to a single unitary as in Methods 1 and 2. 
Due to the larger added number of variational degrees of freedom, Method 3 produces a greater reduction in the McLachlan distance $\mathcal L^2$ per iteration as compared to Methods 1 and 2.
In practice, we find that Methods 2 and 3 produce circuits of similar depth. 
We therefore adopt Method 3 for all following AVQDS(T) simulations. In practice, we choose $\mathcal L_\textrm{cut}^2=10^{-3}$ for noiseless simulations and $\mathcal L_\textrm{cut}^2=10^{-1}$ for noisy simulations.

\section{Models}
\label{sec: model}
Three prototypical one-dimensional (1D) spin models are used to benchmark the AVQDS(T) method: the transverse field Ising model (TFIM), the mixed field Ising model (MFIM), and the Heisenberg Model (HM). 
We consider sudden quench protocols described by time-dependent Hamiltonians of the form
$\hat{H}(t)=\hat{H}_0+\hat{H}_1\Theta(t)$, where $\Theta(t)$ is the Heaviside step function. 
The reference state for the variational ansatz is the initial state $\ket{\Psi_0}=\ket{\Psi(t=0)}$, which is usually chosen to be the ground state of $H_0$ here. 

The Hamiltonian for the MFIM reads as
\be
 \h_{\text{MFIM}} = -J\sum^{N_q}_{i=1} \hat{Z}_i \hat{Z}_{i+1} + \sum^{N_q}_{i=1} \left( h_x\, \hat{X}_i + h_z\, \hat{Z}_i\right),
\ee
where $\hat{X}$, $\hat{Y}$ and $\hat{Z}$ are Pauli operators. 
We consider periodic boundary conditions (PBC): $\hat{\sigma}_{N_q+1}=\hat{\sigma}_1$ with $\sigma=X,Y$ or $Z$. 
We set $J=1$, $h_x=-2$, and consider two different values of $h_z$ ($h_z=0, 0.5$) in our simulations. 
When $h_z=0$, the MFIM is reduced to the TFIM. 
The initial Hamiltonian for $t=0$ is set to be the ferromagnetic Ising model $H_0=-J\sum^{N_q}_{i=1} \hat{Z}_i \hat{Z}_{i+1}$, and the mixed fields are turned on at $t=0$.
Initially, the system is prepared in the all spin-up state, $\ket{\Psi_0}=\ket{\up \dots \up}$, which is a ground state of the ferromagnetic Ising model. 

The Hamiltonian for the HM is given by
\be
 \h_{\text{HM}} = J\sum^{N_q}_{i=1} \left(\hat{X}_i \hat{X}_{i+1} + \hat{Y}_i \hat{Y}_{i+1} + \hat{Z}_i \hat{Z}_{i+1}\right),
\ee
where we assume PBC. 
Again, $H_0$ is set to be the simple Ising model $H_0=J\sum^{N_q}_{i=1} \hat{Z}_i \hat{Z}_{i+1}$. 
Here, we consider the antiferromagnetic coupling ($J=1$) because for the ferromagnetic case, the ground state of the Ising model remains the ground state of the HM. 
For the antiferromagnetic HM, we prepare the system in the N\'{e}el state $\ket{\Psi_0}=\lvert\uparrow,\downarrow,\uparrow,\downarrow,\cdots\rangle$ with perfect antiferromagnetic ordering at $t=0$; 
quantum spin dynamics occurs 
as $\ket{\Psi_0}$ is not an eigenstate of $\h_{\text{HM}}$. 

In addition to AVQDS and AVQDS(T), we also simulated quantum dynamics with the first-order Trotter decomposition~\cite{trotter, nielsen2002quantum} and VQDS with a fixed ansatz, namely the Hamiltonian variational ansatz (HVA)~\cite{wecker2015_trotterizedsp}. 
In Trotter decomposition, the time evolution of a quantum state over a single time step $\delta t$ is approximated as $\ket{\Psi(t+\delta t)} \approx \prod\limits_{\mu}e^{-i\delta tj_\mu \, \hat{h}_\mu}\ket{\Psi(t)}$, where $j_\mu\hat{h}_\mu$ are the terms in the Hamiltonian: $\h=\sum\limits_\mu j_\mu\hat{h}_\mu$. 
$\hat{h}_\mu$ are usually Pauli strings and $j_\mu$ are constants.
A basic feature of such Trotter circuits is that the circuit depth keeps increasing with the simulation time $t$ as the circuit depth scales linearly with $t$.  
Motivated by the form of these Trotter circuits, the HVA entails parameterized unitary gates generated by the individual Pauli terms $\hat{h}_\mu$. 
The ansatz uses a fixed number of layers, $L$, and the rotation angles entering the parameterized unitaries are allowed to vary in space and from layer to layer. 
Thus, the variational quantum state within HVA can be written as
\be
\ket{\Psi[\bth]}=\prod_{s=1}^{L}\left[\prod_\mu\exp(-i\theta^s_\mu \hat{h}_\mu) \right]\ket{\Psi_0},
\ee
in which the rotation angles $\theta^s_\mu$ are the variational parameters. In practice, it is desirable to group the Pauli terms $\hat{h}_\mu$ into a number of sub-layers, so that the resulting quantum circuit is more compact. 
For example, in the TFIM, all the Pauli strings in the Hamiltonian are grouped into three sub-layers in a ``brick-wall" fashion: $\{Z_{2i-1}Z_{2i}, 1\leq i\leq N_q/2\}$, $\{Z_{2i}Z_{2i+1}, 1\leq i\leq N_q/2\}$, and $\{X_i, 1\leq i\leq N_q\}$.

\section{Results and Discussions}
\label{sec: results}

\subsection{Benchmarking AVQDS(T) on noiseless simulators}

\begin{figure}[!ht]
	\centering
	\includegraphics[width=0.6\columnwidth]{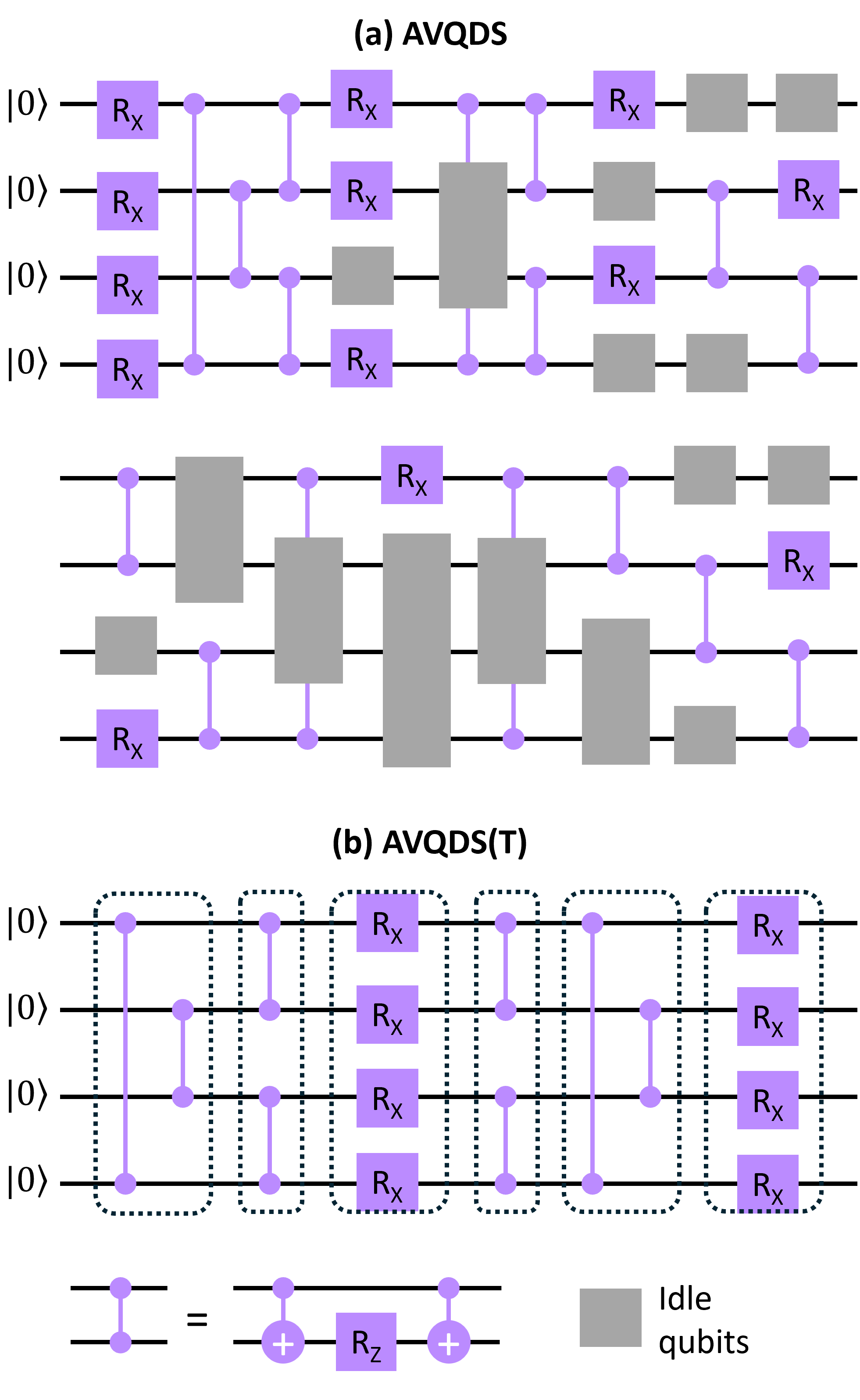}
	\caption{
	The final circuit generated in (a) AVQDS and (b) AVQDS(T) for quench dynamics simulations of the TFIM with $N_q=4$. The AVQDS circuit has a total depth of 17 with idle qubits on many layers. On the other hand, the AVQDS(T) circuit can be separated into 6 complete layers.
	}
	\label{circuit}
\end{figure}
In Fig.~\ref{circuit}, we show quantum circuits generated by AVQDS and AVQDS(T) for quench dynamics simulations of the TFIM with $N_q=4$. 
The AVQDS circuit has a total depth of 17 with idle qubits (grey boxes) in most of the layers. 
On the other hand, the circuit generated in AVQDS(T) is much more compact with only 6 layers. 
The AVQDS(T) circuit is the same as the two-layer HVA ansatz, except that the order of two-qubit gates is switched between adjacent layers.
\begin{figure}[!ht]
	\centering
	\includegraphics[width=\columnwidth]{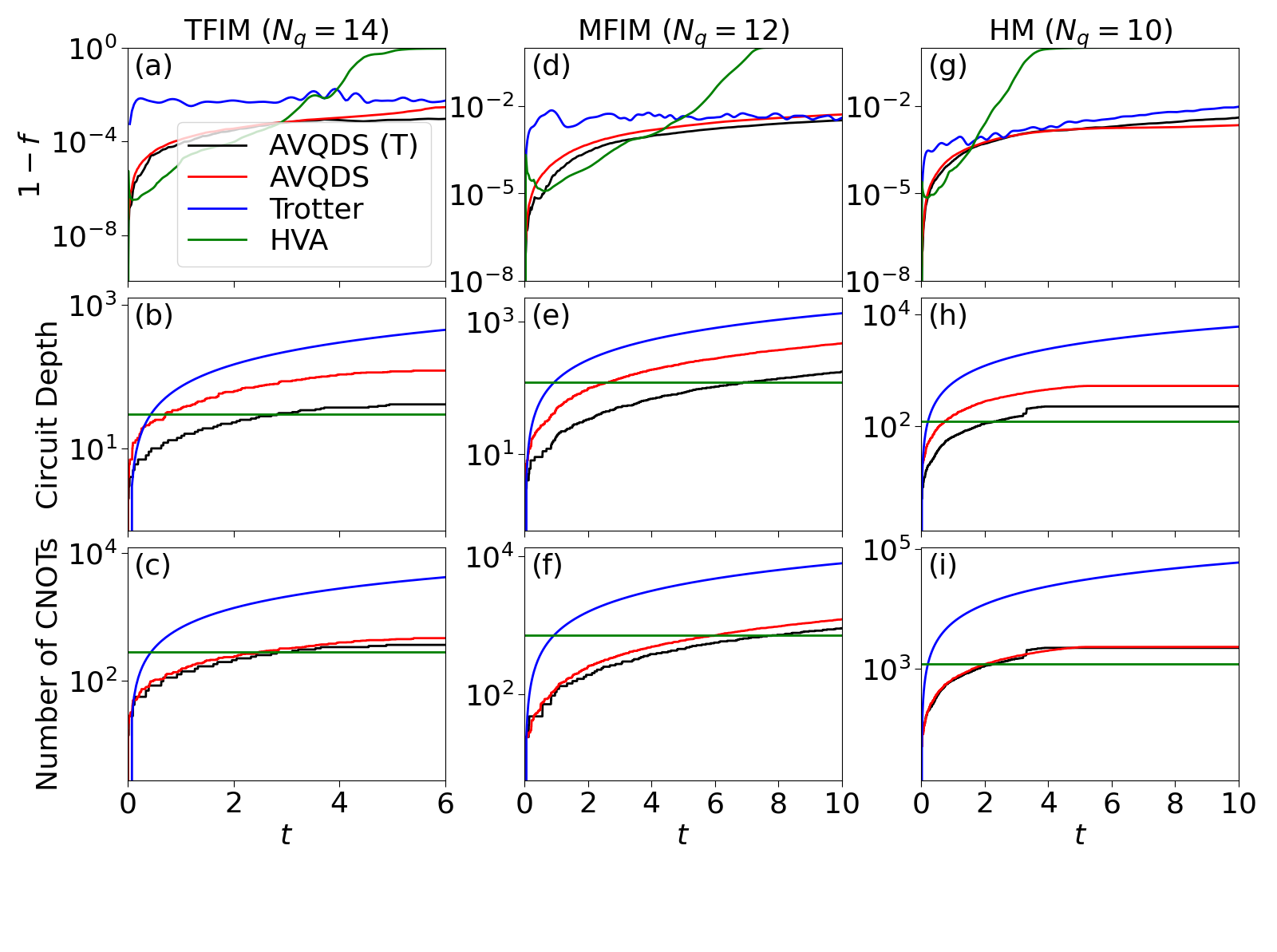}
        \vspace{-1.1cm}
	\caption{
	Quench dynamics of the TFIM ($N_q=14$), MFIM ($N_q=12$) and HM ($N_q=10$) simulated using the AVQDS(T), original AVQDS, Trotter decomposition, and HVA methods. We use $J=1$, $h_x = -2$, and $h_z = 0.5$ for the MFIM and an initial state as described in the text. The time step $\delta t$ used in the Trotter decomposition is set to $0.04$, $0.03$, and $0.01$ for the TFIM, MFIM, and HM, respectively. For HVA, the number of layers $L=10$, $30$, and $20$ for the TFIM, MFIM, and HM, respectively. (a), (d) and (g) show the evolution of the wavefunction infidelity; (b), (e) and (h) show the the evolution of the quantum circuit depth; and (c), (f) and (i) show the evolution of the number of CNOT gates during the simulations.
	}
	\label{benchmark}
\end{figure}

In the following, we benchmark the AVQDS(T) method against the original AVQDS, as well as the first-order Trotter decomposition and VQDS with HVA on noiseless simulators.
We choose $N_q=14$, 12, and 10 for the TFIM, MFIM, and HM, respectively. 
The time step $\delta t$ used in Trotterization is set to 0.04, 0.03, and 0.01, for the TFIM, MFIM, and HM, respectively. 
These time steps are chosen so that the Trotter decomposition method produces comparable accuracy to the AVQDS and AVQDS(T) methods. 
For HVA, we set the layer number $L=10$, 30, 20, corresponding to fixed circuit depths of 30, 120, and 120 for the TFIM, MFIM, and HM, respectively. Here, circuit depth is defined by the number of gate layers in the circuit. 
Here we do not consider optimizations of specific circuit implementations for simplicity.

In Fig.~\ref{benchmark}, we plot the evolution of the wavefunction infidelity $1-f$, the circuit depth, and the number of CNOT gates during quench dynamics simulations of the three models. 
The AVQDS and AVQDS(T) methods generate similar wavefunction infidelities, both smaller than 1\%, for all three models (see top panel of Fig.~\ref{benchmark}).
On the other hand, for all three systems the circuit depth at the end of the AVQDS(T) simulations is reduced by more than half relative to AVQDS (see middle panel of Fig.~\ref{benchmark}). This reduction is important as the circuit depth determines the total execution time of the circuit on hardware, which is upper bounded by the coherence time of the device. Compressing the circuits thus allows reaching a larger final simulation time on hardware.
While the total number of unitaries in the final circuit is similar for AVQDS(T) and AVQDS, the number of resource-intensive two-qubit gates is reduced by 22\% and 26\% for the TFIM ($N_q=14$) and the MFIM ($N_q=12$), respectively, resulting in a noticeable reduction in the number of CNOT gates, as can be seen in Fig.~\ref{benchmark} (c) and (f). However, we believe this reduction is problem-dependent, and can hardly be generalized to an arbitrary Hamiltonian. 
Such an effect is inapplicable to the HM ($N_q=10$) since we choose to only include two-qubit operators in HM simulations. 

In contrast, if the standard Trotter decomposition technique is used to perform the dynamical simulations, the circuit depth and the number of CNOTs must increase by over an order of magnitude compared with AVQDS(T) in order to reach similar accuracy.  
HVA exemplifies a group of methods with a fixed quantum circuit depth. 
As shown in the top panel of Fig.~\ref{benchmark}, HVA can deliver the required accuracy in the early stages of the simulations. 
However, after a certain period of time, the applied HVA circuits completely fail to describe the quantum dynamics as the wavefunction infidelity increases rapidly and approaches one. 
Deeper circuits would have been required at this late stage. 
This behavior, observed in all the models we have studied, demonstrates a typical disadvantage of using a fixed-depth ansatz for quantum-dynamical simulations: one either uses a deep ansatz for the entire duration of the simulation (even when it is not needed to represent the time evolved state at early times) or loses accuracy at long times.
On the other hand, with the ability to expand the ansatz on demand, AVQDS(T) represents a much more efficient strategy for allocating quantum resources. 

\subsection{Implementation of realistic noise channels}

In practical quantum computing, NISQ hardware is subject to various error sources including coherent errors caused by imperfect gate operations and stochastic errors due to qubit decoherence, dephasing, and relaxation. Here we investigate how these hardware imperfections affect quantum dynamical simulations using the TFIM with $N_q=6$ as an example. The final converged ansatz obtained in a noiseless simulation is used in these noisy simulations. We implement a phenomenological noise model proposed by Kandala {\it et al}.~\cite{hardware_efficient_vqe}, which consists of an amplitude damping channel ($\Lambda_a[\rho] = \sum_{i=1}^2 E_i^a \rho E_i^{a\dagger}$) and a dephasing channel ($\Lambda_d[\rho] = \sum_{i=1}^2 E_i^d \rho E_i^{d\dagger}$). Here, $\rho$ is the qubit density matrix. The Kraus operators are defined as follows:
\bea
E_1^a &=& 
\begin{pmatrix}
1 & 0\\
0 & \sqrt{1-p^a}
\end{pmatrix}, 
E_2^a = 
\begin{pmatrix}
0 & \sqrt{p^a}\\
0 & 0
\end{pmatrix}, \nonumber \\
E_1^d &=& 
\begin{pmatrix}
1 & 0\\
0 & \sqrt{1-p^d}
\end{pmatrix}, 
E_2^d = 
\begin{pmatrix}
0 & 0\\
0 & \sqrt{p^d}
\end{pmatrix}.
\eea
The error rates $p^a = 1-e^{-t_g/T_1}$ and $p^d = 1-e^{-2t_g/T_{\phi}}$ depend on the gate time $t_g$, the qubit relaxation time $T_1$, and the dephasing time $T_{\phi}=2T_1T_2/(2T_1-T_2)$, where $T_2$ is the qubit coherence time. For simplicity, we use a uniform single-qubit gate error rate $p_1^a = p_1^d \equiv p_1 = 10^{-4}$, which is typical of the current hardware. The two-qubit gate error is also assumed to be uniform  $p_2^a = p_2^d \equiv p_2$, with $10^{-4}<p_2<10^{-2}$~\cite{barison2022variational, mukherjee2023comparative}. We used $2^{14}$ shots for all measurements. The calculations were performed using the QASM simulator provided in the quantum computing package Qiskit~\cite{Qiskit}.

\begin{figure}[!ht]
	\centering
	\includegraphics[width=0.9\columnwidth]{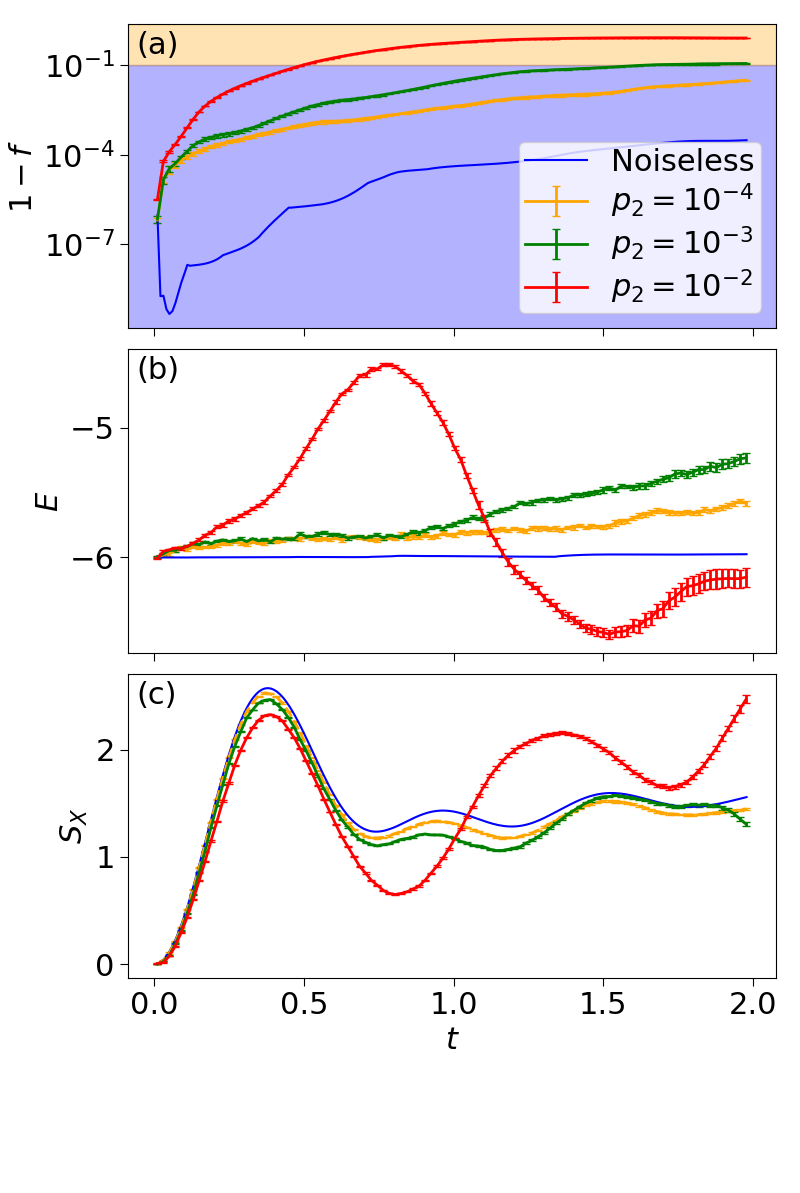}
        \vspace{-1.4cm}
	\caption{
	Quench dynamics simulations of the TFIM ($N_q=6$) using the QASM simulator of Qiskit with various two-qubit noise rates. The final saturated ansatz constructed during a noiseless simulation was used for the simulation. The noisy results are averaged over 20 independent runs. Noiseless results are also shown for comparison. (a), (b), and (c) show the time evolution of the wave function infidelity, the measured total energy, and the$X$-component of the magnetization $S_X\equiv\bra{\Psi(t)}\sum_{i=1}^{N_q}X_i\ket{\Psi(t)}$, respectively.  The background colors in (a) separate the acceptance range with $f>0.9$.
	}
	\label{n6-error}
\end{figure}

In Fig.~\ref{n6-error}, we show the results of the noisy simulations with the two-qubit error rate $p_2=10^{-4},10^{-3}$, and $10^{-2}$, together with the results obtained on a noiseless simulator. The wave function fidelity progressively worsens as $p_2$ increases compared with the noiseless results, as can be seen in Fig.~\ref{n6-error} (a). The light purple region in Fig.~\ref{n6-error} (a) shows an ``acceptance range" with $f>0.9$. For $p_2=10^{-4}$ and $10^{-3}$, the fidelity is largely within this range in the entire duration of $t<2$, while the simulation with $p_2=10^{-2}$ leaves the acceptance range at $t\sim 0.5$. The measured total energy is plotted in Fig.~\ref{n6-error} (b). For this closed system, the exact total energy is kept constant at $E=-6$, which is accurately captured in the noiseless simulation. For $p_2=10^{-4}$ and $10^{-3}$, the measured total energy gradually deviates from the exact value during the simulation, with the error growing to 7\% and 12\% at $t=2$, respectively. For $p_2=10^{-2}$, the total energy oscillates around the exact value with the largest deviation exceeding 20\%. We also measured the $X$-component of the magnetization defined as $S_X=\bra{\Psi(t)}\sum_{i=1}^{N_q}X_i\ket{\Psi(t)}$, as shown in Fig.~\ref{n6-error} (c). For both $p_2=10^{-4}$ and $p_2=10^{-3}$, the trajectory of $S_X$ generally follows that of the noiseless simulation, with $p_2=10^{-4}$ giving an overall better agreement. For $p_2=10^{-2}$, on the other hand, $S_X$ starts to develop a large deviation from the noiseless values at $t=0.6$.

\subsection{Comparison with the adaptive pVQD method}
The projected variational quantum dynamics (pVQD) method~\cite{Barison2021efficientquantum} and its adaptive version~\cite{Linteau2024adaptiveprojectedVQD} represent an alternative approach for simulating quantum dynamics using adaptively parameterized quantum circuits. In pVQD, the variational parameters $\bth$ are updated at each time step by maximizing the overlap between the parametrized state at the next time step $\ket{\Psi[\bth+\boldsymbol{\delta\theta}]}$ and the Trotter-evolved state from the current parameters $\hat{\mathcal{T}}(\delta t)\ket{\Psi[\bth]}$, where the Trotter operator $\hat{\mathcal{T}}$ is an approximation of the exact time evolution operator $\hat{\mathcal{T}}(\delta t)\approx e^{-i\hat{H}\delta t}$, and $\delta t$ is a predetermined time-step size. As pointed out in Ref.~\cite{Barison2021efficientquantum}, in the limit of $\delta t\rightarrow 0$, pVQD is equivalent to the McLachlan variational principle implemented in AVQDS(T).

It will bring useful insight into the general performance of variational quantum algorithms in simulating quantum dynamics by comparing these two methods. This comparison naturally incorporates the impact of time-step size. For simplicity, we assume a constant number of shots for each distinct circuit used to measure quantities of interest. For example, the same number of shots is applied to measure a quantum state overlap in adaptive pVQD and an element of the quantum geometric tensor in AVQDS. This assumption allows us to focus solely on the number of distinct circuits for comparison.

In AVQDS~\cite{AVQDS}, the number of distinct circuits for measuring $M$, $V$ and $L^2$ at each time step can be estimated as $N_H(N_{\bth}-1)+N_{\bth}(N_{\bth}+1)/2+N_{H,H^2}^c$, where $N_{\bth}$ is the number of variational parameters, $N_H$ is the number of Pauli strings in the Hamiltonian $\hat{H}$, which scales linearly with $N_q$ for the local spin models studied in this paper. The constant $N_{H,H^2}^c$ is the number of partitions of the commuting Pauli strings in $\hat{H}$ and $\hat{H}^2$. If the ansatz adaptive process is triggered, it requires additional  $N_H N_{\bth}$ circuits for scanning the operator pool composed of the $N_H$ non-identity Pauli strings in the Hamiltonian. For the adaptive pVQD method, the dominant computational cost is the measurement of gradients using the parameter shift rule during the scan of the pool and the optimization of the cost function. Accordingly, the number of distinct circuits at each time step can be estimated as  $2N_{\bth}^0 N_\mathrm{iter}^0+\sum_{i=1}^{N_\mathrm{adapt}}(2N_H+2N_{\bth}^iN_\mathrm{iter}^i)$, where $N_\mathrm{adapt}$ is the total number of adaptive steps in the current time step (which can be zero), $N_{\bth}^i$ and $N_\mathrm{iter}^i$ are the number of parameters and the number of iterations for optimizing the cost function after the $i$th adaptive step, respectively. 

To illustrate this comparison concretely, we analyze quench dynamics simulations of the transverse-field Ising model (TFIM) with $N_q=10$ sites. The adaptive pVQD calculations were performed using the open-source code~\cite{pvqd_github} with the default settings for trotterization (second order with 2 repetitions), which allows for relatively large time-step sizes.  Specifically, $\delta t=0.16$ gives the most accurate result for simulations run up to $t=2$, with a minimal final infidelity $1-f=0.012$. $7.78\times10^5$ distinct circuits are required for quantum measurements during the simulation. 
For AVQDS, a much smaller time-step size of 0.02 is necessary to achieve similar infidelity of $1-f=0.010$ at $t=2$. However, the cumulative total number of distinct circuits is $7.31\times10^5$, which is comparable to the adaptive pVQD method with $\delta t=0.16$. The final circuit depth for the variational ansatz in adaptive pVQD and AVQDS simulations is 22 for both methods. For AVQDS, the most complicated circuit is represented by measuring the element of the quantum geometric tensor involving the final ansatz parameter, which introduces an ancillary qubit and enlarges the circuit depth to $22+4=26$, due to the controlled two-qubit Pauli gates. While in adaptive pVQD, one needs to double the ansatz circuit plus an extra trotter circuit (second order with two repetitions) for measuring the cost function, which makes the circuit depth $22\times2+(4\times3-3)=53$ (each Trotter layer has a depth of 3 for the TFIM).

\section{Leveraging classical computation in AVQDS(T)} 
\label{sec: qpu-tn}
For practical AVQDS(T) calculations on QPUs, most quantum resource intensive component is the measurement overhead for the $M_{\mu\nu}$ elements, which scales quadratically with the number of variational parameters $N_{\bth}$~\cite{AVQDS}. 
Since $M_{\mu\nu}$ is equivalent to the real part of the quantum geometric tensor (QGT) or quantum Fisher information matrix~\cite{meyer2021fisher} and therefore has wide applications in quantum science, methods to efficiently measure it on QPUs are under active development~\cite{meyer2021fisher, kolotouros2024acceleratingquantumimaginarytimeevolution, Liu2020QuantumFI, Gacon2021simultaneousPS}, primarily utilizing stochastic approaches~\cite{meyer2021fisher, kolotouros2024acceleratingquantumimaginarytimeevolution}.
Here we propose a complementary approach to save quantum resources without sacrificing accuracy by maximally leveraging classical computational resources thanks to the algorithmic structure of AVQDS(T). 
Specifically, one can capitalize on the observation that the measurement of an element $M_{\mu\nu}$ ($\mu\leq\nu$) only invokes the ansatz circuit fragment $\hat{U}_{\nu}=\prod_{\gamma=1}^{\nu}e^{-i\theta_\gamma\hat{\A}_\gamma}$ up to a depth defined by the $\nu$th parameter, rather than the full ansatz circuit containing $N_{\bth}$ parameters~\cite{AVQDS}. 
This implies that an inner block of the matrix $M$ can always be measured using low-depth circuits, amenable to efficient classical computation with high accuracy. 
In particular, tensor network based techniques can be naturally exploited to boost the efficiency of AVQDS(T) calculations. 
For dynamics simulations starting with easily prepared states, such as product states as demonstrated in this work, the AVQDS(T) circuit depth gradually grows from $\mathcal{O}(1)$ as time evolves. 
This allows the AVQDS(T) simulations for very early times to be executed accurately using only classical resources.
For instance, matrix product states can efficiently simulate quantum circuits of arbitrary width with depths up to around 10, assuming the entangling gates are applied between neighboring qubits, such as in one-dimensional local spin models. This limitation arises because the bond dimension, \(\chi\), increases with each layer of local entangling gates, with the growth being upper-bounded by a factor of 2. 
QPUs become necessary only when the AVQDS(T) circuit depth surpasses \(\sim \log_2(\chi)\) at a certain simulation time \(t\). Beyond this threshold, QPUs are required to measure the remaining \(M_{\mu\nu}\) elements, which lie beyond the capabilities of classical methods.

\begin{figure}[!ht]
	\centering
	\includegraphics[width=1\columnwidth]{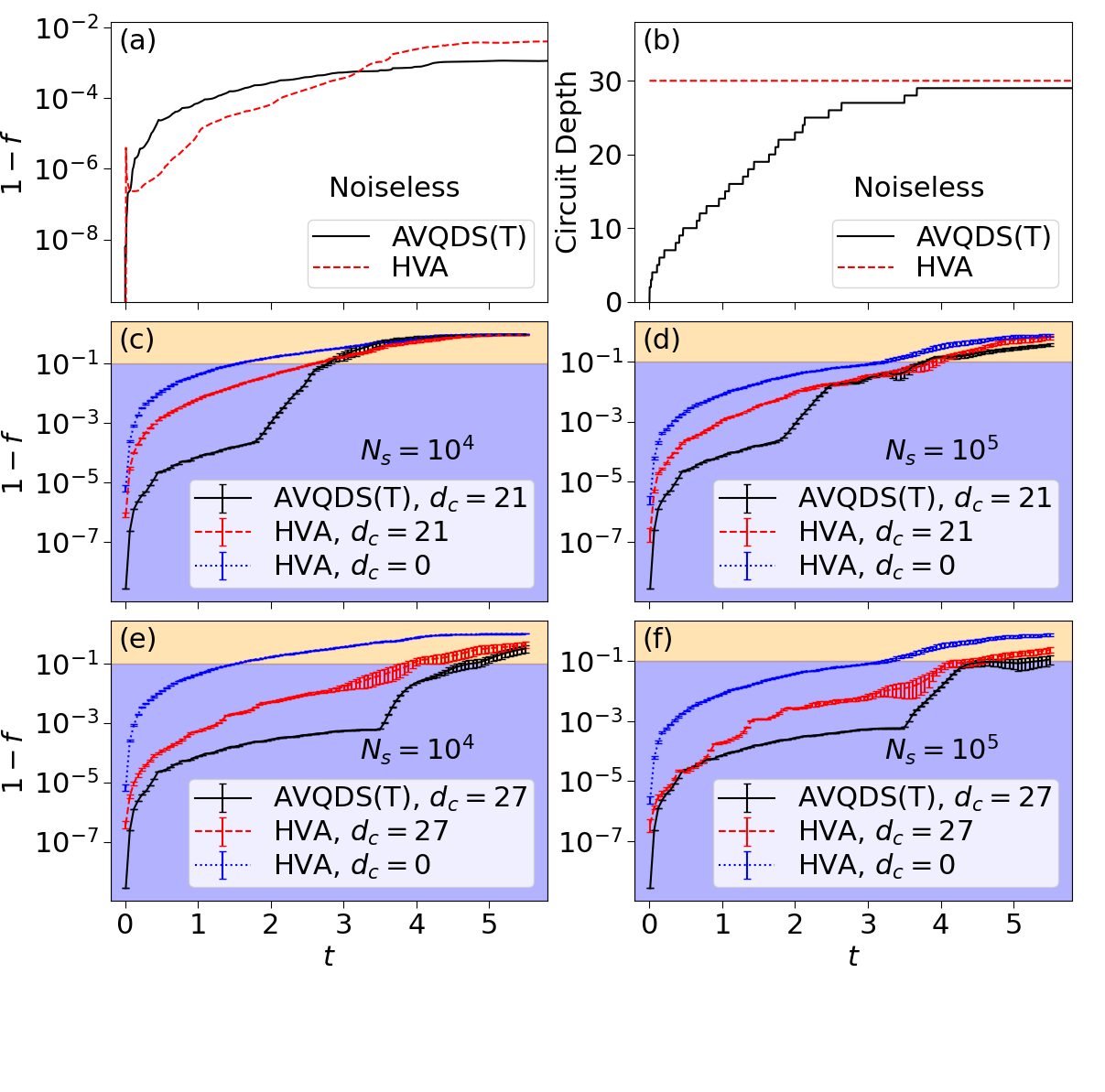}
        \vspace{-1.5cm}
	\caption{
	Quench dynamics simulations for the TFIM with $J=1$, $h_x = -2$ and $N_q=10$ using AVQDS(T) and HVA. HVA with $L=10$ layers is adopted, which amounts to a fixed circuit depth of 30. (a) and (b) show the time evolution of the wavefunction infidelity and the circuit depth of a noiseless simulation, respectively; (c)-(f) show the time evolution of the infidelity on noisy simulators with various $d_c$ and $N_s$. The error bars in (c)-(f) are based on 20 simulations performed independently.
	}
	\label{noise}
\end{figure}

As a proof-of-principle demonstration of the proposed approach, we apply AVQDS(T) to quench dynamics simulations of the TFIM with $N_q=10$ on a simulator including shot noise. 
An inner block of the $M$ matrix up to a dimension set by a threshold circuit depth $d_c$ is accurately evaluated using a classical algorithm, namely exact diagonalization for simplicity. 
The remaining $M_{\mu\nu}$ elements are measured with  $N_s$ ``shots" in the manner described in Appendix~\ref{appendix:solvers}, mimicking the noisy matrix elements that would be obtained on a QPU.
(Recall that shot noise is incorporated as Gaussian random noise with standard deviation determined by $N_s$, which can also be viewed as a proxy for other types of noise one would encounter on a QPU.)
Since the focus here is on the efficacy of partial measurement of the $M$ matrix, we set the $N_{\bth}$ elements of $V$~\eqref{eq: V} to their state vector values (i.e., those that would be obtained with $N_s=\infty$). 
We contrast the performance of AVQDS(T) with that of HVA with $L=10$ layers. 
The HVA circuit amounts to a depth of $30$, which is comparable to that of the final AVQDS(T) circuit. 
In contrast to AVQDS(T), the $M$ matrix for VQDS with HVA has a fixed dimension and only the inner block of dimension $d_c (< 30)$ is evaluated ``classically" (i.e., without noise). 
Therefore, shot noise impacts the entire fixed-ansatz VQDS simulation from the very beginning, unlike in AVQDS(T) where it only becomes important when the adaptive ansatz depth exceeds $d_c$.

To contextualize the noisy simulations, we start with the results obtained on a noiseless simulator.  
The time trace of the wavefunction infidelity in Fig.~\ref{noise} (a) demonstrates that both AVQDS(T) and HVA can accurately describe the dynamics for the entire duration of the simulation, with infidelity smaller than $10^{-2}$. 
The black curve in Fig.~\ref{noise} (b) shows the growth of the circuit depth in AVQDS(T), which gives a saturated depth of 29, close to the fixed value of 30 for the HVA circuit. 

To demonstrate the impact of noise in AVQDS(T), $d_c$ needs to be less than the saturated circuit depth; we choose $d_c=21$ and $27$. 
We also set an upper bound of $30$ on the ansatz circuit depth for AVQDS(T), matching that of the fixed-depth HVA circuit. 
In Fig.~\ref{noise} (c)-(f), we plot the wavefunction infidelity as a function of time for the noisy simulations. 
The time traces are averaged over 20 independent runs for each parameter set, with the error bars showing the standard deviation. 
The light purple regions denote the acceptance range with $f>0.90$. 
AVQDS(T) simulations are noiseless for times corresponding to circuit depths $d\leq d_c$, leading to more accurate results than HVA in the early stage of the simulation, even though the AVQDS(T) circuit is much shallower. 
This is in contrast with the noiseless case shown in Fig.~\ref{noise} (a), where the HVA wavefunction, in general, has a higher fidelity when $t$ is small. 
After the noise kicks in when $d>d_c$, the AVQDS(T) wavefunction infidelity increases more rapidly, creating a kink in the trajectory. This demonstrates that noise has a significant detrimental effect on the dynamics simulation fidelity even when only a small fraction of the components of $M$ are noisy. 
For both $d_c=21$ and $27$, the wavefunctions generated by AVQDS(T) remain more accurate than HVA for a certain time period. 
Moreover, when $d_c$ is increased to $27$, simulations with AVQDS (T) appear to stay in the acceptance range noticeably longer than HVA, as shown in Fig.~\ref{noise} (e) and (f). 
While the time traces generated by the two methods eventually merge, these results demonstrate that with the ability to leverage classical algorithms, dynamical simulations with AVQDS(T) can achieve more accurate results for a considerable time duration by deferring the onset of noise. Even more importantly, by adaptively expanding the ansatz over time and leveraging classical simulations until $d > d_c$, we shift the use of quantum resources to where they are most needed. Quantum resources are only used beyond classical simulation times and only to simulate those circuits that are cannot be executed on classical computers. 
For reference, VQDS results with HVA obtained exclusively on noisy simulators by setting $d_c=0$, are also included in Fig.~\ref{noise} (c)-(f), where one can see that even for a fixed ansatz, the dynamical simulations can still benefit greatly from accurately evaluating an inner section of the $M$ matrix using a classical algorithm. 

Finally, we compare the number of distinct circuits required for quantum measurements to demonstrate the effectiveness of the hybrid classical/quantum strategy in saving quantum resources. Using $d_c=21$ and $N_s=10^4$ as an example [see Fig.~\ref{noise}(c)], for $t<2.7$ when the fidelity is within the acceptance range, the required total number of distinct circuits for AVQDS is $2.0\times10^6$, while the number of distinct circuits for HVA with the same circuit-depth cutoff during the same period of time is nearly an order of magnitude larger: $1.5\times10^7$. Furthermore, if HVA is run exclusively on noisy simulators ($d_c=0$), the number of distinct circuits is $3.6\times10^7$, which is more than doubled compared with HVA that implements classical algorithms for $d\leq21$.

\section{Conclusion}
\label{sec: conclusion}
We report progress on enabling adaptive variational quantum dynamics simulations on noisy quantum hardware with the goal to reach beyond-classical simulation times. Specifically, we discuss three improvements compared to the original AVQDS method~\cite{AVQDS}: (i) we implement a TETRIS approach to significantly compress the circuit depth. 
Unlike the original AVQDS, in which the operators are added to the variational ansatz one at a time, AVQDS(T) adds a series of operators acting on disjoint sets of qubits at each iteration. 
(ii) We benchmark a noise-resilient scheme to solve the dynamical equations of motion for the variational parameters based on a truncation of the eigenvalues of the matrix $M$ (the real part of the QGT), and (iii) we leverage classical computing resources by computing a part of $M$ on classical hardware and use quantum resources only during beyond-classical timescales and circuit depths. We believe that this synergistic interplay between classical and quantum computing is one of the main advantages of variational approaches to quantum dynamics simulations, where one can defer the use of quantum resources to the beyond-classical regime and employ them when they are most effective. 

To benchmark these advancements, we demonstrated AVQDS(T) calculations for the TFIM, MFIM and HM with $N_q\geq10$. 
In addition to greatly reducing the circuit depth compared with AVQDS and Trotter decomposition, AVQDS(T) also generates circuits with fewer 2-qubit unitaries, better suited for current and near-term QPUs. The circuit compression is relevant to reach longer simulation times as the circuit depth sets the execution time on hardware, which is upper bounded by the device's coherence time. In a comparative study, we demonstrate that the required quantum resources are comparable in AVQDS(T) and an alternative adaptive pVQD method in order to achieve similar accuracies, while the overall circuit depth in AVQDS(T) is significantly shallower than the latter. Simulations implementing realistic noise channels show that under a typical error rate of $10^{-4}$ for single-qubit gates, the quantum observables agree reasonably well with exact values for small error rates for two-qubit gates ($10^{-4}$ and $10^{-3}$), while showing qualitative differences with exact values for larger two-qubit error rates.
Finally, we propose a way to save quantum resources in AVQDS(T) calculations by exploiting classical algorithms to reduce the quantum measurement overhead of the quantum Fisher information matrix $M$, which is the most demanding part of the algorithm. 
Classical computation can be leveraged in the whole simulation period because an inner block of $M$ involving low-depth ansatz circuit fragments can always be efficiently evaluated. 
The gradual circuit growth entailed by the AVQDS(T) algorithm also renders it feasible to simulate the dynamics at very early times using only classical computations. 
With a proof-of-principle demonstration including shot noise effects, we show that AVQDS(T) aided with classical evaluations of a sub-block of $M$ produces more accurate dynamics than HVA for a notable time duration at comparable circuit depths. 
In view of the fact that classical algorithms like TN-based approaches, both rigorous and approximate, are being actively developed for quantum circuit simulations~\cite{Ayral2023, Tindall2024EfficientTN}, we envision this synergistic engagement of both quantum and classical resources, in particular the partial classical evaluation of the $M$-matrix, will help boost AVQDS(T) simulations to large-size systems for showcasing quantum utility~\cite{herrmann2023quantumud, kimEvidenceUtilityQuantum2023}.

\section*{Acknowledgements}
This work was supported by the U.S. Department of Energy (DOE), Office of Science, Basic Energy Sciences, Materials Science and Engineering Division. The research was performed at the Ames National Laboratory, which is operated for the U.S. DOE by Iowa State University under Contract No. DE-AC02-07CH11358. The authors acknowledge J. Aftergood for generating scripts for drawing the quantum circuits.

\newpage
\appendix
\section{Solubility of the Equations of Motion}
\label{appendix:a}
When the $N_{\bth} \times N_{\bth}$ real symmetric matrix $M$ is singular, there exists a non-zero real vector $x\in\mathbb{R}^{N_{\bth}}$ satisfying $Mx=0$. In order for Eq.~\eqref{eq: eom} $M\dot{\bth}=V$ to remain solvable, $x$ needs to be orthogonal to $V$ ($x^\dagger V=\sum_\mu x_\mu V_\mu=0$). To see this, assume there is a solution $\dot{\bth_0}$ satisfying $M\dot{\bth_0}=V$. $x^\dagger M=0$ because $M$ is real symmetric. Then, $x^\dagger V=x^\dagger M\dot{\bth_0}=0$. In the following, we show that this condition is satisfied for the $M$ and $V$ defined in Eqs.~\eqref{eq: M} and~\eqref{eq: V}, respectively.

We define $\ket{\xi_\mu}\equiv\frac{\partial \ket{\Psi}}{\partial \theta_\mu}$ with normalized $\braket{\Psi|\Psi} = 1$. Then, $M_{\mu\nu}$ and $V_\mu$ can be written as 
\begin{align}
    M_{\mu \nu} &= \Re\left[\ov{\xi_\mu}{\xi_\nu} - \ov{\xi_\mu}{\Psi} \ov{\Psi} {\xi_\nu}\right] \label{eq: newM} \\
    V_\mu &= \Im\left[\bra{\xi_\mu}\h\ket{\Psi} - \ov{\xi_\mu}{\Psi}\Av{\Psi}{\h} \right]\,. \label{eq: newV}
\end{align}
 
We first show that $M$ is positive semi-definite. Indeed, for any $y\in\mathbb{R}^{N_{\bth}}$, 
\bea
y^\dagger My &=& \sum_{\mu\nu}y_\mu M_{\mu\nu}y_\nu\notag \\
&=& \sum_{\mu\nu}\Re\left[y_\mu\ov{\xi_\mu}{\xi_\nu}y_\nu - y_\mu\ov{\xi_\mu}{\Psi} \ov{\Psi}{\xi_\nu}y_\nu\right] \notag \\
&=& \olp{\chi} - \ov{\chi}{\Psi}\ov{\Psi}{\chi} \notag \\
&\geq & 0, \label{eq: posidef}
\eea
where we define $\ket{\chi}\equiv \sum_\mu y_\mu\ket{\xi_\mu}$. 
The equal sign in Eq.~\eqref{eq: posidef} is reached if and only if $\ket{\chi} = \alpha\ket{\Psi}$ with $\alpha$ being a constant ($\alpha$ can be zero). The positive semi-definiteness of $M$ ensures that $Mx=0\Leftrightarrow x^\dagger Mx=0$. In other words, the null space of $M$ is formed by vectors $x$ satisfying $\sum_\mu x_\mu\ket{\xi_\mu}=\alpha\ket{\Psi}$. For any $x$ satisfying this condition,
\bea
\sum_\mu x_\mu V_\mu&=&\Im\left[\sum_\mu x_\mu\bra{\xi_\mu}\h\ket{\Psi} + x_\mu \braket{\Psi|\xi_\mu}\Av{\Psi}{\h} \right] \notag \\
&=&\Im\left[(\alpha+\alpha^*)\Av{\Psi}{\h}\right] \notag \\
&=&0. \notag
\eea

\section{Technical approaches for solving the equations of motion}
\label{appendix:solvers}
\begin{figure}[!ht]
	\centering
	\includegraphics[width=\columnwidth]{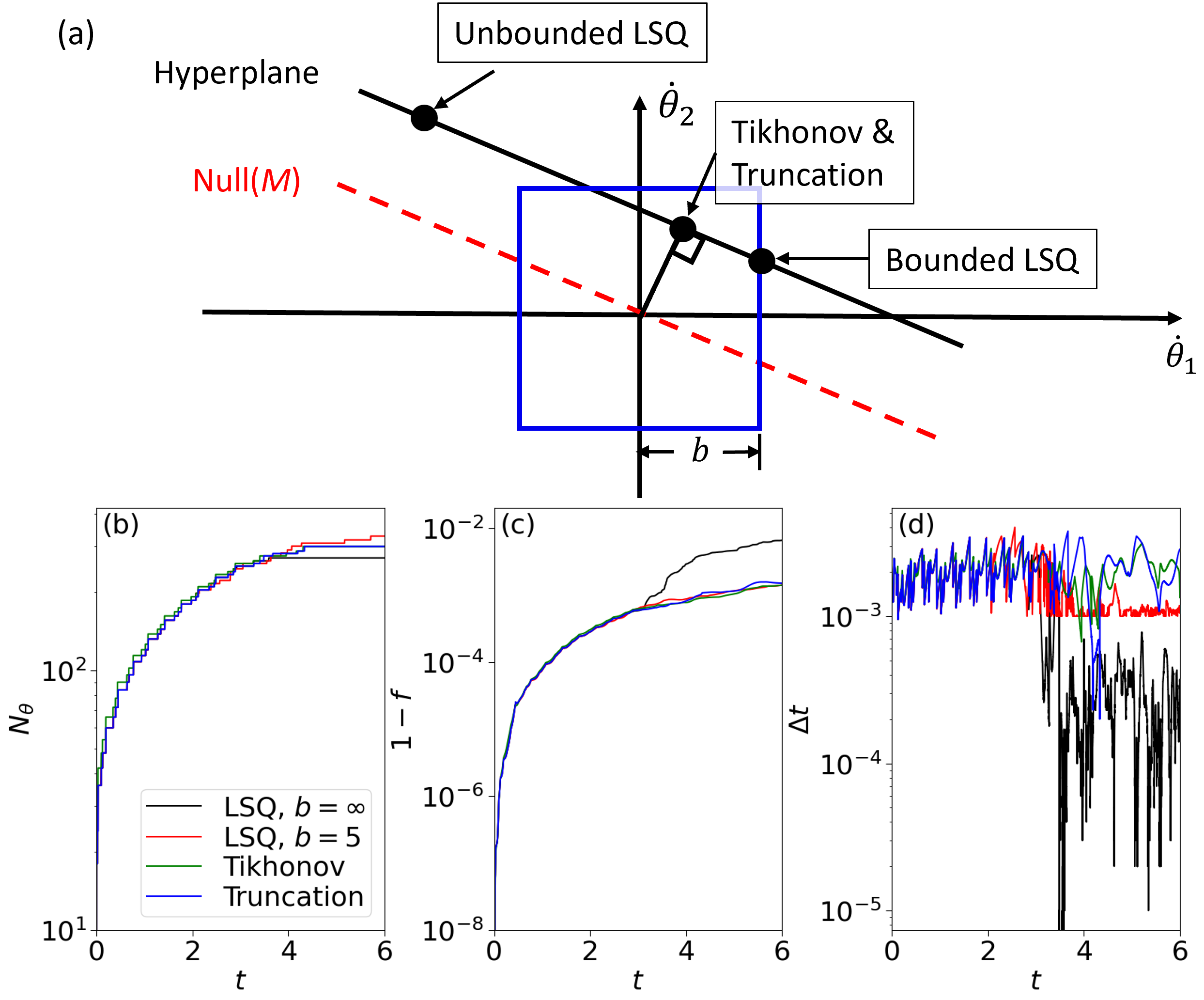}
        \vspace{-0.5cm}
	\caption{
	(a) Schematic showing different approaches for solving Eq.~\eqref{eq: eom} in a 2D parameter space. The red dashed line shows the null space of $M$ (the collection of all $x$ satisfying $Mx=0$). The solid black line parallel to $\text{Null}(M)$ is the hyperplane containing all the solutions of Eq.~\eqref{eq: eom}. The black circles show the solutions that can be obtained using different methods. In particular, the Tikhonov regularization and the truncation methods can locate the unique solution orthogonal to $\text{Null}(M)$. (b) The growth of the number of unitaries $N_{\bth}$ as a function of time, (c) the time evolution of the wavefunction infidelity, and (d) the time increment $\Delta t=\Delta\theta_m/\max_\mu\abs{\dot{\theta}_\mu}$ at each time step as a function of time, during the quench dynamics simulation of the TFIM with $N_q=12$, using the methods outlined in (a) to solve Eq.~\eqref{eq: eom}.
	}
	\label{eom-solver}
\end{figure}

An essential process in both VQDS and AVQDS is to solve the equations of motion Eq.~\eqref{eq: eom}, which can face a numerical complication due to the singularity of the matrix $M$ as defined in Eq.~\eqref{eq: M}; that is, there exist non-zero vectors $x$ that satisfy $Mx=0$. 
In Appendix~\ref{appendix:a}, we show that Eq.~\eqref{eq: eom} remains solvable since the vector $V$ given in Eq.~\eqref{eq: V} is always orthogonal to the null space of $M$. 
In this scenario, Eq.~\eqref{eq: eom} has infinitely many solutions that lie on a hyperplane in the $N_{\bth}$-dimensional parameter space. 
We show in Fig.~\ref{eom-solver} (a) a schematic of the hyperplane in a 2D space, which reduces to a straight line. 
Obviously one cannot solve Eq.~\eqref{eq: eom} by directly inverting $M$. 
Instead, one could try to minimize $\norm{M\dot{\bth}-V}^2$ using least-square (LSQ) techniques~\cite{endo2020calculation}. 
An unbounded LSQ search can in principle reach any solution on the hyperplane, which can have a large magnitude. 
Consequently, a tiny time increment $\Delta t$ must be taken so that the changes in the rotation angles $\Delta\theta_\mu=\dot{\theta}_\mu\Delta t$ remain small enough to guarantee a continuous integration of the equations of motion. 
A remedy for this problem is to set a bound $b$ for the LSQ method, so the search is restricted to be within a hypercube of side $2b$ centered at the origin. 
An exact solution can still be found as long as the hypercube intersects with the hyperplane [see Fig.~\ref{eom-solver} (a)]. 

Tikhonov regularization~\cite{hanke1989regularization} is another popular method for solving linear systems of equations of the form~\eqref{eq: eom} with singular or ill-conditioned $M$.
In this method, the solution to Eq.~\eqref{eq: eom} is computed as $\dot\bth=(M+\varepsilon I)^{-1}V$, where $\varepsilon$ is a small number and $I$ is the identity matrix. 
To better understand the effect of Tikhonov regularization, we first diagonalize the symmetric matrix $M$ with a unitary transformation $M=U\Lambda U^\dagger$, where $\Lambda$ is a diagonal matrix containing the eigenvalues $\lambda_\mu$ of $M$ in ascending order and where each column of $U$ gives the corresponding eigenvector.
The eigenvalues $\lambda_\mu\geq0$ since $M$ is positive semi-definite (see Appendix~\ref{appendix:a}). 
Assume the null space of $M$ has dimension $N_\text{null}$; then, $\lambda_\mu=0$ for $1\leq\mu\leq N_\text{null}$, and the null space of $M$ is spanned by the first $N_\text{null}$ eigenvectors of $M$: $\text{Null}(M)=\text{Span}\{u_1,u_2,\cdots,u_{N_\text{null}}\}$ where $u_\mu$ is the $\mu^\text{th}$ column of $U$. 
Under Tikhonov regularization, $\dot\bth=(M+\varepsilon I)^{-1}V=U(\Lambda+\varepsilon I)^{-1}U^\dagger V$.
$(\Lambda+\varepsilon I)^{-1}$ is also a diagonal matrix dominated by the first $N_\text{null}$ diagonal elements ($\varepsilon^{-1}$). 
However, since $V\perp\text{Null}(M)$, the first $N_\text{null}$ elements of $U^\dagger V$ are zero. Thus, as long as $\varepsilon$ remains large compared to the machine accuracy, the first $N_\text{null}$ elements of the resulting vector $(\Lambda+\varepsilon I)^{-1}U^\dagger V$, which gives the projection of $\dot\bth$ onto $\text{Null}(M)$, are essentially zero. 
Meanwhile, the Tikhonov parameter $\varepsilon$ should also be much smaller than the nonzero eigenvalues $\lambda_\mu > 0$. Thus,
$\dot\bth$ calculated in this way is close to the unique solution of Eq.~\eqref{eq: eom} that is orthogonal to $\text{Null}(M)$.
Such a solution is also schematically shown in Fig.~\ref{eom-solver} (a). 

Alternatively, the coefficient matrix $M$ can be regularized by performing a truncation on its eigenvalues~\cite{hansen1990trucated}.
After diagonalizing $M$, Eq.~\eqref{eq: eom} can be rewritten as $\Lambda (U^\dagger\dot{\bth})=U^\dagger V$. 
Since the first $N_\text{null}$ elements of both $\lambda_\mu$ and $U^\dagger V$ are zero, a solution of the above linear system of equations can be straightforwardly obtained by letting $(U^\dagger\dot{\bth})_\mu=0$ for $1\leq\mu\leq N_\text{null}$, and $(U^\dagger\dot{\bth})_\mu=(U^\dagger V)_\mu/\lambda_\mu$ for $N_\text{null}<\mu\leq N_{\bth}$. 
In other words, $U^\dagger\dot{\bth}$ is an $N_{\bth}$-dimensional vector whose first $N_\text{null}$ elements are zeros.
After $U^\dagger\dot{\bth}$ is obtained, $\dot{\bth}$ is readily available as $\dot{\bth}=U(U^\dagger\dot{\bth})$. This procedure again leads to the unique solution located in the subspace orthogonal to $\text{Null}(M)$ as shown in Fig.~\ref{eom-solver} (a). 
In practice, $N_\text{null}$ is identified by truncating the eigenvalues based on a threshold value $\varepsilon$; in other words, $N_\text{null}=\left|\{\lambda_\mu|\lambda_\mu\leq\varepsilon\}\right|$ where $|\cdot|$ denotes the size of a set. 
On noiseless simulators, this method can be interpreted as a modified Tikhonov regularization, where the uniform shift of the eigenvalues of $M$ by $\varepsilon I$ is replaced with a partial shift involving only the negligible eigenvalues within the null space. 
However, when noise is present, the first $N_\text{null}$ eigenvalues of $M=M_0+\delta M$ will take nonzero values $\delta\lambda_\mu$. 
Here, $M_0$ and $\delta M$ denote the exact coefficient matrix and the noise, respectively. 
At the same time, $u_\mu^\dagger V$ ($\mu\leq N_\text{null}$) will not vanish either, and their values $\delta(u_\mu^\dagger V)$ are also determined by the noise.
Consequently, the projection of $\dot{\bth}$ onto $\text{Null}(M_0)$, calculated within Tikhonov regularization via $(\delta\lambda_i+\varepsilon)^{-1}\delta(u_i^\dagger V)$, can be large because it involves division of small quantities. 
On the other hand, such operations are largely avoided in the truncation method by effectively truncating the eigenvalues $\delta\lambda_\mu$ that are only nonzero due to noise, resulting in a solution in the vicinity of the orthogonal subspace of $\text{Null}(M_0)$. This method, which we call the ``truncation" method, has been applied to variational quantum imaginary time evolution~\cite{gacon2023ieee}. Here, we examine its performance on simulating real-time quantum dynamics on both noiseless and noisy simulators.

We compare the above-mentioned methods for solving the equations of motion Eq.~\eqref{eq: eom} by performing AVQDS(T) on the TFIM with $N_q=12$. We here use method 3 of Fig.~\ref{adgp} to construct the compressed circuits. 
$\varepsilon$ is set to $10^{-6}$ for both the Tikhonov regularization and truncation methods. 
For the bounded LSQ search, we set the bound $b=5$.
The variational parameters $\bth$ are updated according to the Euler method $\Delta\bth=\dot{\bth}\Delta t$ with $\Delta t=\Delta\theta_m/\max_\mu\abs{\dot{\theta}_\mu}$~\cite{iserles2009ode}, where the maximal allowed change in rotation angles $\Delta\theta_m$ is set to 0.005.
Fig.~\ref{eom-solver} (b) plots the growth of the number of unitaries $N_{\bth}$ as a function of $t$, which shows that the different methods produce circuits of similar complexity.
We show the infidelity $1-f$ of $\ket{\Psi[\bth(t)]}$ and the time increment at each time step ($\Delta t$) as a function of $t$ in Fig.~\ref{eom-solver} (c) and (d), respectively. 
The fidelity $f$ is defined as $f \equiv \abs{\ov{\Psi[\bth(t)]}{\Psi_\text{exact}(t)}}^2$, where $\ket{\Psi_\text{exact}(t)}=e^{-i\h t}\ket{\Psi_0}$ is obtained by directly applying the time-evolution operator.
All the curves coincide for $t<2$ in both Fig.~\ref{eom-solver} (c) and (d), indicating the matrix $M$ is well-conditioned, and all the methods can effectively locate the unique solution to Eq.~\eqref{eq: eom} during this time period.  

For the unbounded LSQ search, $\Delta t$ drops by $\sim 2$ orders of magnitude when $t>3$, as can be observed in Fig.~\ref{eom-solver} (d), indicating that the method finds solutions that have large magnitude with the matrix $M$ being singular. 
This drastically slows down the simulation; and, even more problematically, increases the error build-up in the wavefunction, resulting in a simultaneous abrupt increase of the infidelity as shown in Fig.~\ref{eom-solver} (c), even though the ansatz generated by the unbounded LSQ search has the lowest number of unitaries during this time period as shown in Fig.~\ref{eom-solver} (a). 
During this period, the bounded LSQ search frequently hits the bound as $\Delta t$ plateaus at $10^{-3}$, which is equal to $\Delta\theta_m/b$. This suggests that the bound $b$ needs to be properly set: if $b$ is too small, it will not be able to reach a solution; while if $b$ is too large, it will slow down the simulation since the allowed $\Delta t$ is inverse-proportional to $b$. 
One can see from Fig.~\ref{eom-solver} (b) that except for the unbounded search, all the other methods deliver similar accuracy, demonstrating they are all effective in mitigating the numerical problem caused by the singularity of $M$ for this specific model. 
We will use the truncation method with $\varepsilon=10^{-6}$ in the following noiseless simulations.

\begin{figure}[!ht]
	\centering
	\includegraphics[width=\columnwidth]{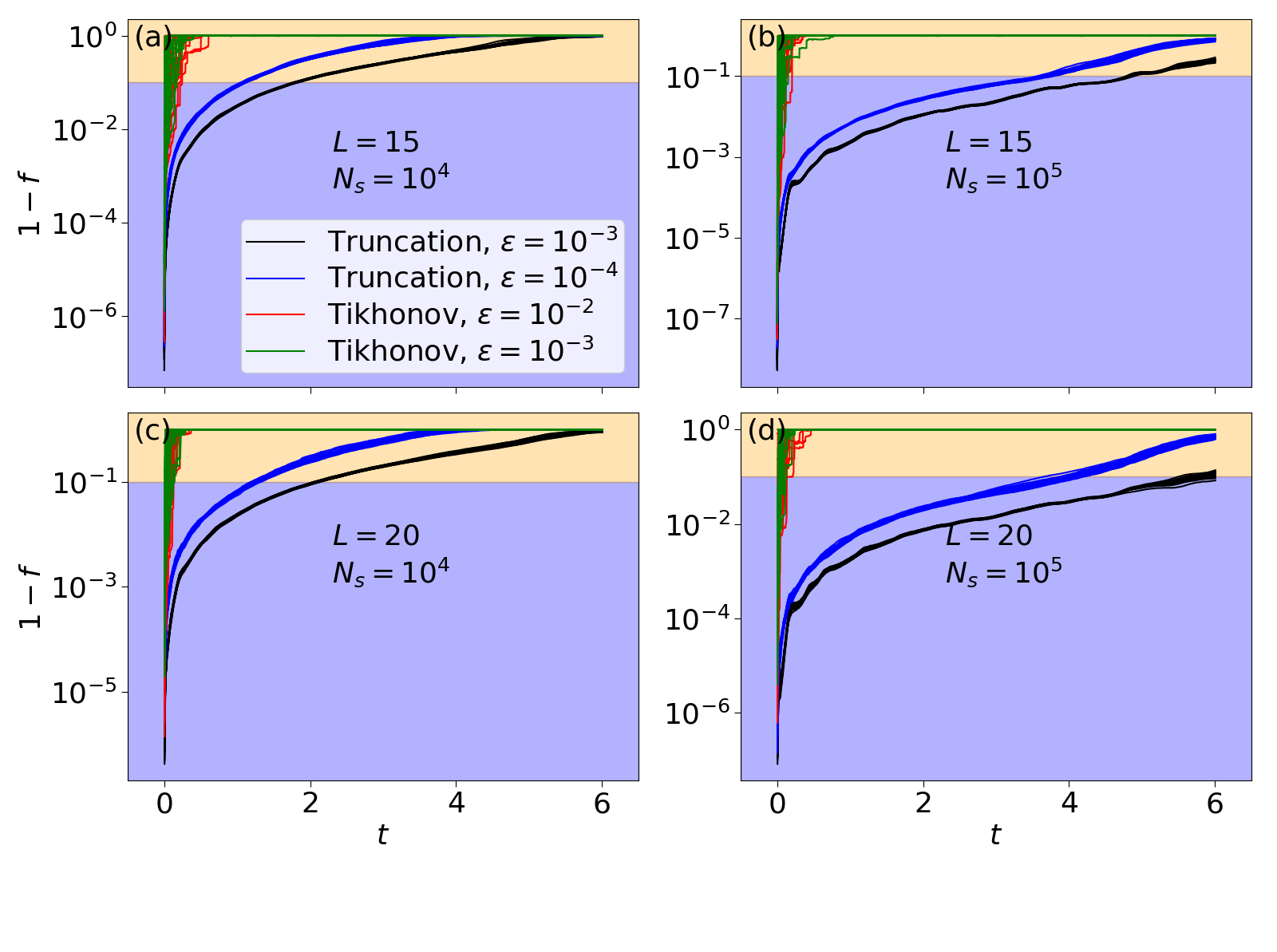}
        \vspace{-1.5cm}
	\caption{
	The evolution of the wavefunction infidelity in the quench dynamics simulation of TFIM with $N_q=10$, using fixed HVA with $L=15$ or 20 layers. Shot noise was added to the matrix $M$ with the number of shots $N_s=10^4$ or $10^5$. (a) to (d) show results for $L=15,N_s=10^4$; $L=15,N_s=10^5$; $L=20,N_s=10^4$; and $L=20,N_s=10^5$, respectively. Two methods were used to regularize $M$: the Tikhonov method with $\varepsilon=10^{-2}$ and $10^{-3}$, and the truncation method with $\varepsilon=10^{-3}$ and $10^{-4}$. Each dataset shown in a different color is an overlay of 20 independent runs. The background colors separate the acceptance range with $f>0.9$. Fixed time steps $\Delta t=0.002$ and $0.005$ were used for $L=15$ and $L=20$, respectively.  
	}
	\label{regularize}
\end{figure}

In the current NISQ era, it is important to examine the noise resilience of different strategies for solving Eq.~\eqref{eq: eom} in a noisy environment. 
Here, we only consider the shot noise on the matrix $M$---resulting from a fixed set of circuit samples or ``shots" performed per matrix element---to single out the effect of these solvers on regularizing $M$. 
The matrix elements of $M$ can be obtained by calculating the probability $p$ that an ancilla qubit measurement yields 1~\cite{AVQDS}, such that $M_{\mu\nu}=(2p-1)/4$.
The variance $\sigma_{M_{\mu\nu}}^2=p(p-1)/(4N_s)$ where $N_s$ is the number of shots used to measure $M_{\mu\nu}$.
We assume that all matrix elements are measured with the same number of shots. 
In our calculations, we simulate the shot noise by replacing the noiseless value $M_{\mu\nu}$ with a random number drawn from a Gaussian distribution with mean $M_{\mu\nu}$ and variance $\sigma_{M_{\mu\nu}}^2$. 

Since the bounded LSQ method requires a predetermined bound that is generally problem-specific, we focus on the Tikhonov regularization and truncation methods, taking the TFIM with $N_q=10$ as an example. 
A fixed HVA is adopted to ensure that the quantum circuits are the same throughout the simulation for both methods.
Relatively deeper circuits are required in the presence of noise; thus, we choose $L=15$ and 20. 
$N_s$ is set to moderate values $10^4$ and $10^5$. 
The parameter $\varepsilon$ controls the small positive value added to the diagonal of the matrix $M$ in the Tikhonov regularization method, or the threshold for truncating the eigenvalues of $M$ in the truncation method.  
Relative to noiseless simulations, a larger $\varepsilon$ is needed so as to avoid being overwhelmed by the noise.
On the other hand, if $\varepsilon$ is too large, it causes too much distortion to the matrix $M$. 
After some pre-experimentation, we choose $\varepsilon=10^{-2}$ and $10^{-3}$ for the Tikhonov regularization method, and $\varepsilon=10^{-3}$ and $10^{-4}$ for the truncation method. 
These parameters yield the same order of magnitude for the standard error of $M_{\mu\nu}$ given $N_s = 10^4 $ and $10^5$. 
We implement a fixed time step $\Delta t$ for this comparison so that the total computation cost is the same for both methods. 
$\Delta t$ is set to 0.002 for $L=15$ and 0.005 for deeper circuits with $L=20$. 
20 independent runs were performed for each parameter set. 
Figure~\ref{regularize} gives the time traces of the wavefunction infidelity, in which the region shaded in light purple shows the ``acceptance range" with $f>0.9$. 
The contrast is clear: the simulations using Tikhonov regularization escape the acceptance range shortly after being launched, while those performed with the truncation method stay within the acceptance range for a much longer time, which increases with the number of HVA layers or the number of shots. 
The failure of the Tikhonov regularization method is due to its difficulty in curbing the components of $\dot{\bth}$ on the null space of the unperturbed coefficient matrix, resulting in large $|\dot{\bth}|$.
One has to reduce $\Delta t$ by about an order of magnitude in order to make it work, greatly increasing the required quantum resources.  
The truncation method with $\varepsilon=10^{-3}$ works the best for all the parameter sets that have been examined, and will be used in subsequent noisy simulations. With this optimal setting, the average number of distinct circuits for the duration within the acceptance range is about $4\times10^{11}$ using $N_s=10^4$ for both $L=15$ and $L=20$. Note that a smaller $\Delta t$ is used for $L=20$. This number increases to $10^{12}$ for $N_s=10^5$ due to the longer duration within the acceptance range. Quantum resources at this level remain challenging in the NISQ era, and manageable jobs on real devices typically feature smaller system sizes or shallower ans{\"a}tze.


\begin{thebibliography}{65}%
\makeatletter
\providecommand \@ifxundefined [1]{%
 \@ifx{#1\undefined}
}%
\providecommand \@ifnum [1]{%
 \ifnum #1\expandafter \@firstoftwo
 \else \expandafter \@secondoftwo
 \fi
}%
\providecommand \@ifx [1]{%
 \ifx #1\expandafter \@firstoftwo
 \else \expandafter \@secondoftwo
 \fi
}%
\providecommand \natexlab [1]{#1}%
\providecommand \enquote  [1]{``#1''}%
\providecommand \bibnamefont  [1]{#1}%
\providecommand \bibfnamefont [1]{#1}%
\providecommand \citenamefont [1]{#1}%
\providecommand \href@noop [0]{\@secondoftwo}%
\providecommand \href [0]{\begingroup \@sanitize@url \@href}%
\providecommand \@href[1]{\@@startlink{#1}\@@href}%
\providecommand \@@href[1]{\endgroup#1\@@endlink}%
\providecommand \@sanitize@url [0]{\catcode `\\12\catcode `\$12\catcode
  `\&12\catcode `\#12\catcode `\^12\catcode `\_12\catcode `\%12\relax}%
\providecommand \@@startlink[1]{}%
\providecommand \@@endlink[0]{}%
\providecommand \url  [0]{\begingroup\@sanitize@url \@url }%
\providecommand \@url [1]{\endgroup\@href {#1}{\urlprefix }}%
\providecommand \urlprefix  [0]{URL }%
\providecommand \Eprint [0]{\href }%
\providecommand \doibase [0]{https://doi.org/}%
\providecommand \selectlanguage [0]{\@gobble}%
\providecommand \bibinfo  [0]{\@secondoftwo}%
\providecommand \bibfield  [0]{\@secondoftwo}%
\providecommand \translation [1]{[#1]}%
\providecommand \BibitemOpen [0]{}%
\providecommand \bibitemStop [0]{}%
\providecommand \bibitemNoStop [0]{.\EOS\space}%
\providecommand \EOS [0]{\spacefactor3000\relax}%
\providecommand \BibitemShut  [1]{\csname bibitem#1\endcsname}%
\let\auto@bib@innerbib\@empty
\bibitem [{\citenamefont {Feynman}(1982)}]{feynman82qc}%
  \BibitemOpen
  \bibfield  {author} {\bibinfo {author} {\bibfnamefont {R.~P.}\ \bibnamefont
  {Feynman}},\ }\bibfield  {title} {\bibinfo {title} {Simulating physics with
  computers},\ }\href {https://doi.org/10.1007/BF02650179} {\bibfield
  {journal} {\bibinfo  {journal} {Int. J. Theor. Phys.}\ }\textbf {\bibinfo
  {volume} {21}},\ \bibinfo {pages} {467} (\bibinfo {year} {1982})}\BibitemShut
  {NoStop}%
\bibitem [{\citenamefont {Lloyd}(1996)}]{lloyd1996}%
  \BibitemOpen
  \bibfield  {author} {\bibinfo {author} {\bibfnamefont {S.}~\bibnamefont
  {Lloyd}},\ }\bibfield  {title} {\bibinfo {title} {Universal quantum
  simulators},\ }\href {https://doi.org/10.1126/science.273.5278.1073}
  {\bibfield  {journal} {\bibinfo  {journal} {Science}\ }\textbf {\bibinfo
  {volume} {273}},\ \bibinfo {pages} {1073} (\bibinfo {year}
  {1996})}\BibitemShut {NoStop}%
\bibitem [{\citenamefont {Abrams}\ and\ \citenamefont
  {Lloyd}(1997)}]{Abrams97simulation}%
  \BibitemOpen
  \bibfield  {author} {\bibinfo {author} {\bibfnamefont {D.~S.}\ \bibnamefont
  {Abrams}}\ and\ \bibinfo {author} {\bibfnamefont {S.}~\bibnamefont {Lloyd}},\
  }\bibfield  {title} {\bibinfo {title} {Simulation of many-body fermi systems
  on a universal quantum computer},\ }\href
  {https://doi.org/10.1103/PhysRevLett.79.2586} {\bibfield  {journal} {\bibinfo
   {journal} {Phys. Rev. Lett.}\ }\textbf {\bibinfo {volume} {79}},\ \bibinfo
  {pages} {2586} (\bibinfo {year} {1997})}\BibitemShut {NoStop}%
\bibitem [{\citenamefont {Abrams}\ and\ \citenamefont
  {Lloyd}(1999)}]{pea_lloyd}%
  \BibitemOpen
  \bibfield  {author} {\bibinfo {author} {\bibfnamefont {D.~S.}\ \bibnamefont
  {Abrams}}\ and\ \bibinfo {author} {\bibfnamefont {S.}~\bibnamefont {Lloyd}},\
  }\bibfield  {title} {\bibinfo {title} {Quantum algorithm providing
  exponential speed increase for finding eigenvalues and eigenvectors},\ }\href
  {https://doi.org/10.1103/PhysRevLett.83.5162} {\bibfield  {journal} {\bibinfo
   {journal} {Phys. Rev. Lett.}\ }\textbf {\bibinfo {volume} {83}},\ \bibinfo
  {pages} {5162} (\bibinfo {year} {1999})}\BibitemShut {NoStop}%
\bibitem [{\citenamefont {Somma}\ \emph {et~al.}(2003)\citenamefont {Somma},
  \citenamefont {Ortiz}, \citenamefont {Knill},\ and\ \citenamefont
  {Gubernatis}}]{somma03quantum}%
  \BibitemOpen
  \bibfield  {author} {\bibinfo {author} {\bibfnamefont {R.}~\bibnamefont
  {Somma}}, \bibinfo {author} {\bibfnamefont {G.}~\bibnamefont {Ortiz}},
  \bibinfo {author} {\bibfnamefont {E.}~\bibnamefont {Knill}},\ and\ \bibinfo
  {author} {\bibfnamefont {J.}~\bibnamefont {Gubernatis}},\ }\bibfield  {title}
  {\bibinfo {title} {Quantum simulations of physics problems},\ }\href@noop {}
  {\bibfield  {journal} {\bibinfo  {journal} {Int. J. Quantum Inf.}\ }\textbf
  {\bibinfo {volume} {1}},\ \bibinfo {pages} {189} (\bibinfo {year}
  {2003})}\BibitemShut {NoStop}%
\bibitem [{\citenamefont {Aspuru-Guzik}\ \emph {et~al.}(2005)\citenamefont
  {Aspuru-Guzik}, \citenamefont {Dutoi}, \citenamefont {Love},\ and\
  \citenamefont {Head-Gordon}}]{asp_ipea}%
  \BibitemOpen
  \bibfield  {author} {\bibinfo {author} {\bibfnamefont {A.}~\bibnamefont
  {Aspuru-Guzik}}, \bibinfo {author} {\bibfnamefont {A.~D.}\ \bibnamefont
  {Dutoi}}, \bibinfo {author} {\bibfnamefont {P.~J.}\ \bibnamefont {Love}},\
  and\ \bibinfo {author} {\bibfnamefont {M.}~\bibnamefont {Head-Gordon}},\
  }\bibfield  {title} {\bibinfo {title} {Simulated quantum computation of
  molecular energies},\ }\href {https://doi.org/10.1126/science.1113479}
  {\bibfield  {journal} {\bibinfo  {journal} {Science}\ }\textbf {\bibinfo
  {volume} {309}},\ \bibinfo {pages} {1704} (\bibinfo {year}
  {2005})}\BibitemShut {NoStop}%
\bibitem [{\citenamefont {Kassal}\ \emph {et~al.}(2008)\citenamefont {Kassal},
  \citenamefont {Jordan}, \citenamefont {Love}, \citenamefont {Mohseni},\ and\
  \citenamefont {Aspuru-Guzik}}]{kassal08polynomial}%
  \BibitemOpen
  \bibfield  {author} {\bibinfo {author} {\bibfnamefont {I.}~\bibnamefont
  {Kassal}}, \bibinfo {author} {\bibfnamefont {S.~P.}\ \bibnamefont {Jordan}},
  \bibinfo {author} {\bibfnamefont {P.~J.}\ \bibnamefont {Love}}, \bibinfo
  {author} {\bibfnamefont {M.}~\bibnamefont {Mohseni}},\ and\ \bibinfo {author}
  {\bibfnamefont {A.}~\bibnamefont {Aspuru-Guzik}},\ }\bibfield  {title}
  {\bibinfo {title} {Polynomial-time quantum algorithm for the simulation of
  chemical dynamics},\ }\href {https://doi.org/10.1073/pnas.0808245105}
  {\bibfield  {journal} {\bibinfo  {journal} {PNAS}\ }\textbf {\bibinfo
  {volume} {105}},\ \bibinfo {pages} {18681} (\bibinfo {year}
  {2008})}\BibitemShut {NoStop}%
\bibitem [{\citenamefont {Georgescu}\ \emph {et~al.}(2014)\citenamefont
  {Georgescu}, \citenamefont {Ashhab},\ and\ \citenamefont {Nori}}]{rmp_qs}%
  \BibitemOpen
  \bibfield  {author} {\bibinfo {author} {\bibfnamefont {I.~M.}\ \bibnamefont
  {Georgescu}}, \bibinfo {author} {\bibfnamefont {S.}~\bibnamefont {Ashhab}},\
  and\ \bibinfo {author} {\bibfnamefont {F.}~\bibnamefont {Nori}},\ }\bibfield
  {title} {\bibinfo {title} {Quantum simulation},\ }\href
  {https://doi.org/10.1103/RevModPhys.86.153} {\bibfield  {journal} {\bibinfo
  {journal} {Rev. Mod. Phys.}\ }\textbf {\bibinfo {volume} {86}},\ \bibinfo
  {pages} {153} (\bibinfo {year} {2014})}\BibitemShut {NoStop}%
\bibitem [{\citenamefont {Wecker}\ \emph
  {et~al.}(2015{\natexlab{a}})\citenamefont {Wecker}, \citenamefont {Hastings},
  \citenamefont {Wiebe}, \citenamefont {Clark}, \citenamefont {Nayak},\ and\
  \citenamefont {Troyer}}]{TroyerQCMB}%
  \BibitemOpen
  \bibfield  {author} {\bibinfo {author} {\bibfnamefont {D.}~\bibnamefont
  {Wecker}}, \bibinfo {author} {\bibfnamefont {M.~B.}\ \bibnamefont
  {Hastings}}, \bibinfo {author} {\bibfnamefont {N.}~\bibnamefont {Wiebe}},
  \bibinfo {author} {\bibfnamefont {B.~K.}\ \bibnamefont {Clark}}, \bibinfo
  {author} {\bibfnamefont {C.}~\bibnamefont {Nayak}},\ and\ \bibinfo {author}
  {\bibfnamefont {M.}~\bibnamefont {Troyer}},\ }\bibfield  {title} {\bibinfo
  {title} {Solving strongly correlated electron models on a quantum computer},\
  }\href {https://doi.org/10.1103/PhysRevA.92.062318} {\bibfield  {journal}
  {\bibinfo  {journal} {Phys. Rev. A}\ }\textbf {\bibinfo {volume} {92}},\
  \bibinfo {pages} {062318} (\bibinfo {year} {2015}{\natexlab{a}})}\BibitemShut
  {NoStop}%
\bibitem [{\citenamefont {Lamm}\ and\ \citenamefont
  {Lawrence}(2018)}]{Trotter_dynamics_Lawrence}%
  \BibitemOpen
  \bibfield  {author} {\bibinfo {author} {\bibfnamefont {H.}~\bibnamefont
  {Lamm}}\ and\ \bibinfo {author} {\bibfnamefont {S.}~\bibnamefont
  {Lawrence}},\ }\bibfield  {title} {\bibinfo {title} {Simulation of
  nonequilibrium dynamics on a quantum computer},\ }\href
  {https://doi.org/10.1103/PhysRevLett.121.170501} {\bibfield  {journal}
  {\bibinfo  {journal} {Phys. Rev. Lett.}\ }\textbf {\bibinfo {volume} {121}},\
  \bibinfo {pages} {170501} (\bibinfo {year} {2018})}\BibitemShut {NoStop}%
\bibitem [{\citenamefont {Smith}\ \emph {et~al.}(2019)\citenamefont {Smith},
  \citenamefont {Kim}, \citenamefont {Pollmann},\ and\ \citenamefont
  {Knolle}}]{Trotter_dynamics_Knolle}%
  \BibitemOpen
  \bibfield  {author} {\bibinfo {author} {\bibfnamefont {A.}~\bibnamefont
  {Smith}}, \bibinfo {author} {\bibfnamefont {M.}~\bibnamefont {Kim}}, \bibinfo
  {author} {\bibfnamefont {F.}~\bibnamefont {Pollmann}},\ and\ \bibinfo
  {author} {\bibfnamefont {J.}~\bibnamefont {Knolle}},\ }\bibfield  {title}
  {\bibinfo {title} {Simulating quantum many-body dynamics on a current digital
  quantum computer},\ }\href {https://doi.org/10.1038/s41534-019-0217-0}
  {\bibfield  {journal} {\bibinfo  {journal} {npj Quantum Inf.}\ }\textbf
  {\bibinfo {volume} {5}},\ \bibinfo {pages} {106} (\bibinfo {year}
  {2019})}\BibitemShut {NoStop}%
\bibitem [{\citenamefont {Chen}\ \emph {et~al.}(2022)\citenamefont {Chen},
  \citenamefont {Burdick}, \citenamefont {Yao}, \citenamefont {Orth},\ and\
  \citenamefont {Iadecola}}]{Chen2022}%
  \BibitemOpen
  \bibfield  {author} {\bibinfo {author} {\bibfnamefont {I.-C.}\ \bibnamefont
  {Chen}}, \bibinfo {author} {\bibfnamefont {B.}~\bibnamefont {Burdick}},
  \bibinfo {author} {\bibfnamefont {Y.-X.}\ \bibnamefont {Yao}}, \bibinfo
  {author} {\bibfnamefont {P.~P.}\ \bibnamefont {Orth}},\ and\ \bibinfo
  {author} {\bibfnamefont {T.}~\bibnamefont {Iadecola}},\ }\bibfield  {title}
  {\bibinfo {title} {Error-mitigated simulation of quantum many-body scars on
  quantum computers with pulse-level control},\ }\href
  {https://doi.org/10.1103/PhysRevResearch.4.043027} {\bibfield  {journal}
  {\bibinfo  {journal} {Phys. Rev. Res.}\ }\textbf {\bibinfo {volume} {4}},\
  \bibinfo {pages} {043027} (\bibinfo {year} {2022})}\BibitemShut {NoStop}%
\bibitem [{\citenamefont {McArdle}\ \emph {et~al.}(2020)\citenamefont
  {McArdle}, \citenamefont {Endo}, \citenamefont {Aspuru-Guzik}, \citenamefont
  {Benjamin},\ and\ \citenamefont {Yuan}}]{rmp_qcc}%
  \BibitemOpen
  \bibfield  {author} {\bibinfo {author} {\bibfnamefont {S.}~\bibnamefont
  {McArdle}}, \bibinfo {author} {\bibfnamefont {S.}~\bibnamefont {Endo}},
  \bibinfo {author} {\bibfnamefont {A.}~\bibnamefont {Aspuru-Guzik}}, \bibinfo
  {author} {\bibfnamefont {S.~C.}\ \bibnamefont {Benjamin}},\ and\ \bibinfo
  {author} {\bibfnamefont {X.}~\bibnamefont {Yuan}},\ }\bibfield  {title}
  {\bibinfo {title} {Quantum computational chemistry},\ }\href
  {https://doi.org/10.1103/RevModPhys.92.015003} {\bibfield  {journal}
  {\bibinfo  {journal} {Rev. Mod. Phys.}\ }\textbf {\bibinfo {volume} {92}},\
  \bibinfo {pages} {015003} (\bibinfo {year} {2020})}\BibitemShut {NoStop}%
\bibitem [{\citenamefont {Miessen}\ \emph {et~al.}(2023)\citenamefont
  {Miessen}, \citenamefont {Ollitrault}, \citenamefont {Tacchino},\ and\
  \citenamefont {Tavernelli}}]{miessen2023quantum}%
  \BibitemOpen
  \bibfield  {author} {\bibinfo {author} {\bibfnamefont {A.}~\bibnamefont
  {Miessen}}, \bibinfo {author} {\bibfnamefont {P.~J.}\ \bibnamefont
  {Ollitrault}}, \bibinfo {author} {\bibfnamefont {F.}~\bibnamefont
  {Tacchino}},\ and\ \bibinfo {author} {\bibfnamefont {I.}~\bibnamefont
  {Tavernelli}},\ }\bibfield  {title} {\bibinfo {title} {Quantum algorithms for
  quantum dynamics},\ }\href {https://doi.org/10.1038/s43588-022-00374-2}
  {\bibfield  {journal} {\bibinfo  {journal} {Nat. Comput. Sci.}\ }\textbf
  {\bibinfo {volume} {3}},\ \bibinfo {pages} {25} (\bibinfo {year}
  {2023})}\BibitemShut {NoStop}%
\bibitem [{\citenamefont {Preskill}(2018)}]{nisq}%
  \BibitemOpen
  \bibfield  {author} {\bibinfo {author} {\bibfnamefont {J.}~\bibnamefont
  {Preskill}},\ }\bibfield  {title} {\bibinfo {title} {Quantum {Computing} in
  the {NISQ} era and beyond},\ }\href
  {https://doi.org/10.22331/q-2018-08-06-79} {\bibfield  {journal} {\bibinfo
  {journal} {Quantum}\ }\textbf {\bibinfo {volume} {2}},\ \bibinfo {pages} {79}
  (\bibinfo {year} {2018})}\BibitemShut {NoStop}%
\bibitem [{\citenamefont {Peruzzo}\ \emph {et~al.}(2014)\citenamefont
  {Peruzzo}, \citenamefont {McClean}, \citenamefont {Shadbolt}, \citenamefont
  {Yung}, \citenamefont {Zhou}, \citenamefont {Love}, \citenamefont
  {{Aspuru-Guzik}},\ and\ \citenamefont
  {O'Brien}}]{peruzzoVariationalEigenvalueSolver2014}%
  \BibitemOpen
  \bibfield  {author} {\bibinfo {author} {\bibfnamefont {A.}~\bibnamefont
  {Peruzzo}}, \bibinfo {author} {\bibfnamefont {J.}~\bibnamefont {McClean}},
  \bibinfo {author} {\bibfnamefont {P.}~\bibnamefont {Shadbolt}}, \bibinfo
  {author} {\bibfnamefont {M.-H.}\ \bibnamefont {Yung}}, \bibinfo {author}
  {\bibfnamefont {X.-Q.}\ \bibnamefont {Zhou}}, \bibinfo {author}
  {\bibfnamefont {P.~J.}\ \bibnamefont {Love}}, \bibinfo {author}
  {\bibfnamefont {A.}~\bibnamefont {{Aspuru-Guzik}}},\ and\ \bibinfo {author}
  {\bibfnamefont {J.~L.}\ \bibnamefont {O'Brien}},\ }\bibfield  {title}
  {\bibinfo {title} {A {{Variational Eigenvalue Solver}} on a {{Photonic
  Quantum Processor}}},\ }\href {https://doi.org/10.1038/ncomms5213} {\bibfield
   {journal} {\bibinfo  {journal} {Nat. Commun.}\ }\textbf {\bibinfo {volume}
  {5}},\ \bibinfo {pages} {1} (\bibinfo {year} {2014})}\BibitemShut {NoStop}%
\bibitem [{\citenamefont {McClean}\ \emph {et~al.}(2016)\citenamefont
  {McClean}, \citenamefont {Romero}, \citenamefont {Babbush},\ and\
  \citenamefont {Aspuru-Guzik}}]{vqe_theory}%
  \BibitemOpen
  \bibfield  {author} {\bibinfo {author} {\bibfnamefont {J.~R.}\ \bibnamefont
  {McClean}}, \bibinfo {author} {\bibfnamefont {J.}~\bibnamefont {Romero}},
  \bibinfo {author} {\bibfnamefont {R.}~\bibnamefont {Babbush}},\ and\ \bibinfo
  {author} {\bibfnamefont {A.}~\bibnamefont {Aspuru-Guzik}},\ }\bibfield
  {title} {\bibinfo {title} {The theory of variational hybrid quantum-classical
  algorithms},\ }\href {https://doi.org/10.1088/1367-2630/18/2/023023}
  {\bibfield  {journal} {\bibinfo  {journal} {New J. Phys.}\ }\textbf {\bibinfo
  {volume} {18}},\ \bibinfo {pages} {023023} (\bibinfo {year}
  {2016})}\BibitemShut {NoStop}%
\bibitem [{\citenamefont {O’Malley}\ \emph {et~al.}(2016)\citenamefont
  {O’Malley}, \citenamefont {Babbush}, \citenamefont {Kivlichan},
  \citenamefont {Romero}, \citenamefont {McClean}, \citenamefont {Barends},
  \citenamefont {Kelly}, \citenamefont {Roushan}, \citenamefont {Tranter},
  \citenamefont {Ding} \emph {et~al.}}]{vqe_pea_h2}%
  \BibitemOpen
  \bibfield  {author} {\bibinfo {author} {\bibfnamefont {P.~J.}\ \bibnamefont
  {O’Malley}}, \bibinfo {author} {\bibfnamefont {R.}~\bibnamefont {Babbush}},
  \bibinfo {author} {\bibfnamefont {I.~D.}\ \bibnamefont {Kivlichan}}, \bibinfo
  {author} {\bibfnamefont {J.}~\bibnamefont {Romero}}, \bibinfo {author}
  {\bibfnamefont {J.~R.}\ \bibnamefont {McClean}}, \bibinfo {author}
  {\bibfnamefont {R.}~\bibnamefont {Barends}}, \bibinfo {author} {\bibfnamefont
  {J.}~\bibnamefont {Kelly}}, \bibinfo {author} {\bibfnamefont
  {P.}~\bibnamefont {Roushan}}, \bibinfo {author} {\bibfnamefont
  {A.}~\bibnamefont {Tranter}}, \bibinfo {author} {\bibfnamefont
  {N.}~\bibnamefont {Ding}}, \emph {et~al.},\ }\bibfield  {title} {\bibinfo
  {title} {Scalable quantum simulation of molecular energies},\ }\href
  {https://doi.org/10.1103/PhysRevX.6.031007} {\bibfield  {journal} {\bibinfo
  {journal} {Phys. Rev. X}\ }\textbf {\bibinfo {volume} {6}},\ \bibinfo {pages}
  {031007} (\bibinfo {year} {2016})}\BibitemShut {NoStop}%
\bibitem [{\citenamefont {Kandala}\ \emph {et~al.}(2017)\citenamefont
  {Kandala}, \citenamefont {Mezzacapo}, \citenamefont {Temme}, \citenamefont
  {Takita}, \citenamefont {Brink}, \citenamefont {Chow},\ and\ \citenamefont
  {Gambetta}}]{hardware_efficient_vqe}%
  \BibitemOpen
  \bibfield  {author} {\bibinfo {author} {\bibfnamefont {A.}~\bibnamefont
  {Kandala}}, \bibinfo {author} {\bibfnamefont {A.}~\bibnamefont {Mezzacapo}},
  \bibinfo {author} {\bibfnamefont {K.}~\bibnamefont {Temme}}, \bibinfo
  {author} {\bibfnamefont {M.}~\bibnamefont {Takita}}, \bibinfo {author}
  {\bibfnamefont {M.}~\bibnamefont {Brink}}, \bibinfo {author} {\bibfnamefont
  {J.~M.}\ \bibnamefont {Chow}},\ and\ \bibinfo {author} {\bibfnamefont
  {J.~M.}\ \bibnamefont {Gambetta}},\ }\bibfield  {title} {\bibinfo {title}
  {Hardware-efficient variational quantum eigensolver for small molecules and
  quantum magnets},\ }\href {https://doi.org/10.1038/nature23879} {\bibfield
  {journal} {\bibinfo  {journal} {Nature}\ }\textbf {\bibinfo {volume} {549}},\
  \bibinfo {pages} {242} (\bibinfo {year} {2017})}\BibitemShut {NoStop}%
\bibitem [{\citenamefont {Ryabinkin}\ \emph {et~al.}(2018)\citenamefont
  {Ryabinkin}, \citenamefont {Yen}, \citenamefont {Genin},\ and\ \citenamefont
  {Izmaylov}}]{qcc_scott2018}%
  \BibitemOpen
  \bibfield  {author} {\bibinfo {author} {\bibfnamefont {I.~G.}\ \bibnamefont
  {Ryabinkin}}, \bibinfo {author} {\bibfnamefont {T.-C.}\ \bibnamefont {Yen}},
  \bibinfo {author} {\bibfnamefont {S.~N.}\ \bibnamefont {Genin}},\ and\
  \bibinfo {author} {\bibfnamefont {A.~F.}\ \bibnamefont {Izmaylov}},\
  }\bibfield  {title} {\bibinfo {title} {Qubit coupled cluster method: a
  systematic approach to quantum chemistry on a quantum computer},\ }\href
  {https://doi.org/10.1021/acs.jctc.8b00932} {\bibfield  {journal} {\bibinfo
  {journal} {J. Chem. Theory Comput.}\ }\textbf {\bibinfo {volume} {14}},\
  \bibinfo {pages} {6317} (\bibinfo {year} {2018})}\BibitemShut {NoStop}%
\bibitem [{\citenamefont {Zhang}\ \emph {et~al.}(2021)\citenamefont {Zhang},
  \citenamefont {Gomes}, \citenamefont {Berthusen}, \citenamefont {Orth},
  \citenamefont {Wang}, \citenamefont {Ho},\ and\ \citenamefont
  {Yao}}]{FengVQE}%
  \BibitemOpen
  \bibfield  {author} {\bibinfo {author} {\bibfnamefont {F.}~\bibnamefont
  {Zhang}}, \bibinfo {author} {\bibfnamefont {N.}~\bibnamefont {Gomes}},
  \bibinfo {author} {\bibfnamefont {N.~F.}\ \bibnamefont {Berthusen}}, \bibinfo
  {author} {\bibfnamefont {P.~P.}\ \bibnamefont {Orth}}, \bibinfo {author}
  {\bibfnamefont {C.-Z.}\ \bibnamefont {Wang}}, \bibinfo {author}
  {\bibfnamefont {K.-M.}\ \bibnamefont {Ho}},\ and\ \bibinfo {author}
  {\bibfnamefont {Y.-X.}\ \bibnamefont {Yao}},\ }\bibfield  {title} {\bibinfo
  {title} {Shallow-circuit variational quantum eigensolver based on
  symmetry-inspired hilbert space partitioning for quantum chemical
  calculations},\ }\href {https://doi.org/10.1103/PhysRevResearch.3.013039}
  {\bibfield  {journal} {\bibinfo  {journal} {Phys. Rev. Res.}\ }\textbf
  {\bibinfo {volume} {3}},\ \bibinfo {pages} {013039} (\bibinfo {year}
  {2021})}\BibitemShut {NoStop}%
\bibitem [{\citenamefont {McClean}\ \emph {et~al.}(2018)\citenamefont
  {McClean}, \citenamefont {Boixo}, \citenamefont {Smelyanskiy}, \citenamefont
  {Babbush},\ and\ \citenamefont {Neven}}]{mcclean2018barren}%
  \BibitemOpen
  \bibfield  {author} {\bibinfo {author} {\bibfnamefont {J.~R.}\ \bibnamefont
  {McClean}}, \bibinfo {author} {\bibfnamefont {S.}~\bibnamefont {Boixo}},
  \bibinfo {author} {\bibfnamefont {V.~N.}\ \bibnamefont {Smelyanskiy}},
  \bibinfo {author} {\bibfnamefont {R.}~\bibnamefont {Babbush}},\ and\ \bibinfo
  {author} {\bibfnamefont {H.}~\bibnamefont {Neven}},\ }\bibfield  {title}
  {\bibinfo {title} {Barren plateaus in quantum neural network training
  landscapes},\ }\href {https://doi.org/10.1038/s41467-018-07090-4} {\bibfield
  {journal} {\bibinfo  {journal} {Nat. Commun.}\ }\textbf {\bibinfo {volume}
  {9}},\ \bibinfo {pages} {4812} (\bibinfo {year} {2018})}\BibitemShut
  {NoStop}%
\bibitem [{\citenamefont {Larocca}\ \emph {et~al.}(2024)\citenamefont
  {Larocca}, \citenamefont {Thanasilp}, \citenamefont {Wang}, \citenamefont
  {Sharma}, \citenamefont {Biamonte}, \citenamefont {Coles}, \citenamefont
  {Cincio}, \citenamefont {McClean}, \citenamefont {Holmes},\ and\
  \citenamefont {Cerezo}}]{Larocca2024}%
  \BibitemOpen
  \bibfield  {author} {\bibinfo {author} {\bibfnamefont {M.}~\bibnamefont
  {Larocca}}, \bibinfo {author} {\bibfnamefont {S.}~\bibnamefont {Thanasilp}},
  \bibinfo {author} {\bibfnamefont {S.}~\bibnamefont {Wang}}, \bibinfo {author}
  {\bibfnamefont {K.}~\bibnamefont {Sharma}}, \bibinfo {author} {\bibfnamefont
  {J.}~\bibnamefont {Biamonte}}, \bibinfo {author} {\bibfnamefont {P.~J.}\
  \bibnamefont {Coles}}, \bibinfo {author} {\bibfnamefont {L.}~\bibnamefont
  {Cincio}}, \bibinfo {author} {\bibfnamefont {J.~R.}\ \bibnamefont {McClean}},
  \bibinfo {author} {\bibfnamefont {Z.}~\bibnamefont {Holmes}},\ and\ \bibinfo
  {author} {\bibfnamefont {M.}~\bibnamefont {Cerezo}},\ }\bibfield  {title}
  {\bibinfo {title} {A review of barren plateaus in variational quantum
  computing},\ }\href {http://arxiv.org/abs/2405.00781} {\bibfield  {journal}
  {\bibinfo  {journal} {arXiv:2405.00781}\ } (\bibinfo {year}
  {2024})}\BibitemShut {NoStop}%
\bibitem [{\citenamefont {Barkoutsos}\ \emph {et~al.}(2018)\citenamefont
  {Barkoutsos}, \citenamefont {Gonthier}, \citenamefont {Sokolov},
  \citenamefont {Moll}, \citenamefont {Salis}, \citenamefont {Fuhrer},
  \citenamefont {Ganzhorn}, \citenamefont {Egger}, \citenamefont {Troyer},
  \citenamefont {Mezzacapo} \emph {et~al.}}]{vqe_uccsd}%
  \BibitemOpen
  \bibfield  {author} {\bibinfo {author} {\bibfnamefont {P.~K.}\ \bibnamefont
  {Barkoutsos}}, \bibinfo {author} {\bibfnamefont {J.~F.}\ \bibnamefont
  {Gonthier}}, \bibinfo {author} {\bibfnamefont {I.}~\bibnamefont {Sokolov}},
  \bibinfo {author} {\bibfnamefont {N.}~\bibnamefont {Moll}}, \bibinfo {author}
  {\bibfnamefont {G.}~\bibnamefont {Salis}}, \bibinfo {author} {\bibfnamefont
  {A.}~\bibnamefont {Fuhrer}}, \bibinfo {author} {\bibfnamefont
  {M.}~\bibnamefont {Ganzhorn}}, \bibinfo {author} {\bibfnamefont {D.~J.}\
  \bibnamefont {Egger}}, \bibinfo {author} {\bibfnamefont {M.}~\bibnamefont
  {Troyer}}, \bibinfo {author} {\bibfnamefont {A.}~\bibnamefont {Mezzacapo}},
  \emph {et~al.},\ }\bibfield  {title} {\bibinfo {title} {Quantum algorithms
  for electronic structure calculations: Particle-hole hamiltonian and
  optimized wave-function expansions},\ }\href
  {https://doi.org/10.1103/PhysRevA.98.022322} {\bibfield  {journal} {\bibinfo
  {journal} {Phys. Rev. A}\ }\textbf {\bibinfo {volume} {98}},\ \bibinfo
  {pages} {022322} (\bibinfo {year} {2018})}\BibitemShut {NoStop}%
\bibitem [{\citenamefont {Grimsley}\ \emph {et~al.}(2019)\citenamefont
  {Grimsley}, \citenamefont {Economou}, \citenamefont {Barnes},\ and\
  \citenamefont {Mayhall}}]{grimsleyAdaptiveVariationalAlgorithm2019}%
  \BibitemOpen
  \bibfield  {author} {\bibinfo {author} {\bibfnamefont {H.~R.}\ \bibnamefont
  {Grimsley}}, \bibinfo {author} {\bibfnamefont {S.~E.}\ \bibnamefont
  {Economou}}, \bibinfo {author} {\bibfnamefont {E.}~\bibnamefont {Barnes}},\
  and\ \bibinfo {author} {\bibfnamefont {N.~J.}\ \bibnamefont {Mayhall}},\
  }\bibfield  {title} {\bibinfo {title} {An adaptive variational algorithm for
  exact molecular simulations on a quantum computer},\ }\href
  {https://doi.org/10.1038/s41467-019-10988-2} {\bibfield  {journal} {\bibinfo
  {journal} {Nat. Commun.}\ }\textbf {\bibinfo {volume} {10}},\ \bibinfo
  {pages} {3007} (\bibinfo {year} {2019})}\BibitemShut {NoStop}%
\bibitem [{\citenamefont {Feniou}\ \emph {et~al.}(2023)\citenamefont {Feniou},
  \citenamefont {Hassan}, \citenamefont {Traor{\'e}}, \citenamefont {Giner},
  \citenamefont {Maday},\ and\ \citenamefont {Piquemal}}]{feniou2023overlapav}%
  \BibitemOpen
  \bibfield  {author} {\bibinfo {author} {\bibfnamefont {C.}~\bibnamefont
  {Feniou}}, \bibinfo {author} {\bibfnamefont {M.}~\bibnamefont {Hassan}},
  \bibinfo {author} {\bibfnamefont {D.}~\bibnamefont {Traor{\'e}}}, \bibinfo
  {author} {\bibfnamefont {E.}~\bibnamefont {Giner}}, \bibinfo {author}
  {\bibfnamefont {Y.}~\bibnamefont {Maday}},\ and\ \bibinfo {author}
  {\bibfnamefont {J.-P.}\ \bibnamefont {Piquemal}},\ }\bibfield  {title}
  {\bibinfo {title} {Overlap-adapt-vqe: practical quantum chemistry on quantum
  computers via overlap-guided compact ans{\"a}tze},\ }\href
  {https://doi.org/10.1038/s42005-023-01312-y} {\bibfield  {journal} {\bibinfo
  {journal} {Communications Physics}\ }\textbf {\bibinfo {volume} {6}},\
  \bibinfo {pages} {192} (\bibinfo {year} {2023})}\BibitemShut {NoStop}%
\bibitem [{\citenamefont {Tang}\ \emph {et~al.}(2021)\citenamefont {Tang},
  \citenamefont {Shkolnikov}, \citenamefont {Barron}, \citenamefont {Grimsley},
  \citenamefont {Mayhall}, \citenamefont {Barnes},\ and\ \citenamefont
  {Economou}}]{MayhallQubitAVQE}%
  \BibitemOpen
  \bibfield  {author} {\bibinfo {author} {\bibfnamefont {H.~L.}\ \bibnamefont
  {Tang}}, \bibinfo {author} {\bibfnamefont {V.}~\bibnamefont {Shkolnikov}},
  \bibinfo {author} {\bibfnamefont {G.~S.}\ \bibnamefont {Barron}}, \bibinfo
  {author} {\bibfnamefont {H.~R.}\ \bibnamefont {Grimsley}}, \bibinfo {author}
  {\bibfnamefont {N.~J.}\ \bibnamefont {Mayhall}}, \bibinfo {author}
  {\bibfnamefont {E.}~\bibnamefont {Barnes}},\ and\ \bibinfo {author}
  {\bibfnamefont {S.~E.}\ \bibnamefont {Economou}},\ }\bibfield  {title}
  {\bibinfo {title} {Qubit-adapt-vqe: An adaptive algorithm for constructing
  hardware-efficient ans\"atze on a quantum processor},\ }\href
  {https://doi.org/10.1103/PRXQuantum.2.020310} {\bibfield  {journal} {\bibinfo
   {journal} {PRX Quantum}\ }\textbf {\bibinfo {volume} {2}},\ \bibinfo {pages}
  {020310} (\bibinfo {year} {2021})}\BibitemShut {NoStop}%
\bibitem [{\citenamefont {Yuan}\ \emph {et~al.}(2019)\citenamefont {Yuan},
  \citenamefont {Endo}, \citenamefont {Zhao}, \citenamefont {Li},\ and\
  \citenamefont {Benjamin}}]{theory_vqs}%
  \BibitemOpen
  \bibfield  {author} {\bibinfo {author} {\bibfnamefont {X.}~\bibnamefont
  {Yuan}}, \bibinfo {author} {\bibfnamefont {S.}~\bibnamefont {Endo}}, \bibinfo
  {author} {\bibfnamefont {Q.}~\bibnamefont {Zhao}}, \bibinfo {author}
  {\bibfnamefont {Y.}~\bibnamefont {Li}},\ and\ \bibinfo {author}
  {\bibfnamefont {S.~C.}\ \bibnamefont {Benjamin}},\ }\bibfield  {title}
  {\bibinfo {title} {Theory of variational quantum simulation},\ }\href
  {https://doi.org/10.22331/q-2019-10-07-191} {\bibfield  {journal} {\bibinfo
  {journal} {Quantum}\ }\textbf {\bibinfo {volume} {3}},\ \bibinfo {pages}
  {191} (\bibinfo {year} {2019})}\BibitemShut {NoStop}%
\bibitem [{\citenamefont {Endo}\ \emph
  {et~al.}(2020{\natexlab{a}})\citenamefont {Endo}, \citenamefont {Sun},
  \citenamefont {Li}, \citenamefont {Benjamin},\ and\ \citenamefont
  {Yuan}}]{Endo20variational}%
  \BibitemOpen
  \bibfield  {author} {\bibinfo {author} {\bibfnamefont {S.}~\bibnamefont
  {Endo}}, \bibinfo {author} {\bibfnamefont {J.}~\bibnamefont {Sun}}, \bibinfo
  {author} {\bibfnamefont {Y.}~\bibnamefont {Li}}, \bibinfo {author}
  {\bibfnamefont {S.~C.}\ \bibnamefont {Benjamin}},\ and\ \bibinfo {author}
  {\bibfnamefont {X.}~\bibnamefont {Yuan}},\ }\bibfield  {title} {\bibinfo
  {title} {Variational quantum simulation of general processes},\ }\href@noop
  {} {\bibfield  {journal} {\bibinfo  {journal} {Phys. Rev. Lett.}\ }\textbf
  {\bibinfo {volume} {125}},\ \bibinfo {pages} {010501} (\bibinfo {year}
  {2020}{\natexlab{a}})}\BibitemShut {NoStop}%
\bibitem [{\citenamefont {Cirstoiu}\ \emph {et~al.}(2020)\citenamefont
  {Cirstoiu}, \citenamefont {Holmes}, \citenamefont {Iosue}, \citenamefont
  {Cincio}, \citenamefont {Coles},\ and\ \citenamefont
  {Sornborger}}]{cirstoiu2020variational}%
  \BibitemOpen
  \bibfield  {author} {\bibinfo {author} {\bibfnamefont {C.}~\bibnamefont
  {Cirstoiu}}, \bibinfo {author} {\bibfnamefont {Z.}~\bibnamefont {Holmes}},
  \bibinfo {author} {\bibfnamefont {J.}~\bibnamefont {Iosue}}, \bibinfo
  {author} {\bibfnamefont {L.}~\bibnamefont {Cincio}}, \bibinfo {author}
  {\bibfnamefont {P.~J.}\ \bibnamefont {Coles}},\ and\ \bibinfo {author}
  {\bibfnamefont {A.}~\bibnamefont {Sornborger}},\ }\bibfield  {title}
  {\bibinfo {title} {Variational fast forwarding for quantum simulation beyond
  the coherence time},\ }\href {https://doi.org/10.1038/s41534-020-00302-0}
  {\bibfield  {journal} {\bibinfo  {journal} {npj Quantum Information}\
  }\textbf {\bibinfo {volume} {6}},\ \bibinfo {pages} {82} (\bibinfo {year}
  {2020})}\BibitemShut {NoStop}%
\bibitem [{\citenamefont {Benedetti}\ \emph {et~al.}(2021)\citenamefont
  {Benedetti}, \citenamefont {Fiorentini},\ and\ \citenamefont
  {Lubasch}}]{benedetti2020hardware}%
  \BibitemOpen
  \bibfield  {author} {\bibinfo {author} {\bibfnamefont {M.}~\bibnamefont
  {Benedetti}}, \bibinfo {author} {\bibfnamefont {M.}~\bibnamefont
  {Fiorentini}},\ and\ \bibinfo {author} {\bibfnamefont {M.}~\bibnamefont
  {Lubasch}},\ }\bibfield  {title} {\bibinfo {title} {Hardware-efficient
  variational quantum algorithms for time evolution},\ }\href
  {https://doi.org/10.1103/PhysRevResearch.3.033083} {\bibfield  {journal}
  {\bibinfo  {journal} {Phys. Rev. Res.}\ }\textbf {\bibinfo {volume} {3}},\
  \bibinfo {pages} {033083} (\bibinfo {year} {2021})}\BibitemShut {NoStop}%
\bibitem [{\citenamefont {Chen}\ \emph {et~al.}(2019)\citenamefont {Chen},
  \citenamefont {Gong}, \citenamefont {Xu}, \citenamefont {Yuan}, \citenamefont
  {Wang}, \citenamefont {Wang}, \citenamefont {Ying}, \citenamefont {Lin},
  \citenamefont {Xu}, \citenamefont {Wu} \emph {et~al.}}]{chen19demonstration}%
  \BibitemOpen
  \bibfield  {author} {\bibinfo {author} {\bibfnamefont {M.-C.}\ \bibnamefont
  {Chen}}, \bibinfo {author} {\bibfnamefont {M.}~\bibnamefont {Gong}}, \bibinfo
  {author} {\bibfnamefont {X.-S.}\ \bibnamefont {Xu}}, \bibinfo {author}
  {\bibfnamefont {X.}~\bibnamefont {Yuan}}, \bibinfo {author} {\bibfnamefont
  {J.-W.}\ \bibnamefont {Wang}}, \bibinfo {author} {\bibfnamefont
  {C.}~\bibnamefont {Wang}}, \bibinfo {author} {\bibfnamefont {C.}~\bibnamefont
  {Ying}}, \bibinfo {author} {\bibfnamefont {J.}~\bibnamefont {Lin}}, \bibinfo
  {author} {\bibfnamefont {Y.}~\bibnamefont {Xu}}, \bibinfo {author}
  {\bibfnamefont {Y.}~\bibnamefont {Wu}}, \emph {et~al.},\ }\bibfield  {title}
  {\bibinfo {title} {Demonstration of adiabatic variational quantum computing
  with a superconducting quantum coprocessor},\ }\href@noop {} {\bibfield
  {journal} {\bibinfo  {journal} {arXiv:1905.03150}\ } (\bibinfo {year}
  {2019})}\BibitemShut {NoStop}%
\bibitem [{\citenamefont {Berthusen}\ \emph {et~al.}(2022)\citenamefont
  {Berthusen}, \citenamefont {Trevisan}, \citenamefont {Iadecola},\ and\
  \citenamefont {Orth}}]{Berthusen:2022}%
  \BibitemOpen
  \bibfield  {author} {\bibinfo {author} {\bibfnamefont {N.~F.}\ \bibnamefont
  {Berthusen}}, \bibinfo {author} {\bibfnamefont {T.~V.}\ \bibnamefont
  {Trevisan}}, \bibinfo {author} {\bibfnamefont {T.}~\bibnamefont {Iadecola}},\
  and\ \bibinfo {author} {\bibfnamefont {P.~P.}\ \bibnamefont {Orth}},\
  }\bibfield  {title} {\bibinfo {title} {Quantum dynamics simulations beyond
  the coherence time on noisy intermediate-scale quantum hardware by
  variational trotter compression},\ }\href
  {https://doi.org/10.1103/PhysRevResearch.4.023097} {\bibfield  {journal}
  {\bibinfo  {journal} {Phys. Rev. Res.}\ }\textbf {\bibinfo {volume} {4}},\
  \bibinfo {pages} {023097} (\bibinfo {year} {2022})}\BibitemShut {NoStop}%
\bibitem [{\citenamefont {Yao}\ \emph {et~al.}(2021)\citenamefont {Yao},
  \citenamefont {Gomes}, \citenamefont {Zhang}, \citenamefont {Wang},
  \citenamefont {Ho}, \citenamefont {Iadecola},\ and\ \citenamefont
  {Orth}}]{AVQDS}%
  \BibitemOpen
  \bibfield  {author} {\bibinfo {author} {\bibfnamefont {Y.-X.}\ \bibnamefont
  {Yao}}, \bibinfo {author} {\bibfnamefont {N.}~\bibnamefont {Gomes}}, \bibinfo
  {author} {\bibfnamefont {F.}~\bibnamefont {Zhang}}, \bibinfo {author}
  {\bibfnamefont {C.-Z.}\ \bibnamefont {Wang}}, \bibinfo {author}
  {\bibfnamefont {K.-M.}\ \bibnamefont {Ho}}, \bibinfo {author} {\bibfnamefont
  {T.}~\bibnamefont {Iadecola}},\ and\ \bibinfo {author} {\bibfnamefont
  {P.~P.}\ \bibnamefont {Orth}},\ }\bibfield  {title} {\bibinfo {title}
  {Adaptive variational quantum dynamics simulations},\ }\href
  {https://doi.org/10.1103/PRXQuantum.2.030307} {\bibfield  {journal} {\bibinfo
   {journal} {PRX Quantum}\ }\textbf {\bibinfo {volume} {2}},\ \bibinfo {pages}
  {030307} (\bibinfo {year} {2021})}\BibitemShut {NoStop}%
\bibitem [{\citenamefont {McLachlan}(1964)}]{mclachlan64variational}%
  \BibitemOpen
  \bibfield  {author} {\bibinfo {author} {\bibfnamefont {A.}~\bibnamefont
  {McLachlan}},\ }\bibfield  {title} {\bibinfo {title} {A variational solution
  of the time-dependent schrodinger equation},\ }\href
  {https://doi.org/10.1080/00268976400100041} {\bibfield  {journal} {\bibinfo
  {journal} {Mol. Phys.}\ }\textbf {\bibinfo {volume} {8}},\ \bibinfo {pages}
  {39} (\bibinfo {year} {1964})}\BibitemShut {NoStop}%
\bibitem [{\citenamefont {Mootz}\ \emph {et~al.}(2024)\citenamefont {Mootz},
  \citenamefont {Orth}, \citenamefont {Huang}, \citenamefont {Luo},
  \citenamefont {Wang},\ and\ \citenamefont {Yao}}]{mootz2023twodimensional}%
  \BibitemOpen
  \bibfield  {author} {\bibinfo {author} {\bibfnamefont {M.}~\bibnamefont
  {Mootz}}, \bibinfo {author} {\bibfnamefont {P.~P.}\ \bibnamefont {Orth}},
  \bibinfo {author} {\bibfnamefont {C.}~\bibnamefont {Huang}}, \bibinfo
  {author} {\bibfnamefont {L.}~\bibnamefont {Luo}}, \bibinfo {author}
  {\bibfnamefont {J.}~\bibnamefont {Wang}},\ and\ \bibinfo {author}
  {\bibfnamefont {Y.-X.}\ \bibnamefont {Yao}},\ }\bibfield  {title} {\bibinfo
  {title} {Two-dimensional coherent spectrum of high-spin models via a quantum
  computing approach},\ }\href {https://doi.org/10.1088/2058-9565/ad57ea}
  {\bibfield  {journal} {\bibinfo  {journal} {Quantum Sci. Technol.}\ }\textbf
  {\bibinfo {volume} {9}},\ \bibinfo {pages} {035054} (\bibinfo {year}
  {2024})}\BibitemShut {NoStop}%
\bibitem [{\citenamefont {Barison}\ \emph {et~al.}(2021)\citenamefont
  {Barison}, \citenamefont {Vicentini},\ and\ \citenamefont
  {Carleo}}]{Barison2021efficientquantum}%
  \BibitemOpen
  \bibfield  {author} {\bibinfo {author} {\bibfnamefont {S.}~\bibnamefont
  {Barison}}, \bibinfo {author} {\bibfnamefont {F.}~\bibnamefont {Vicentini}},\
  and\ \bibinfo {author} {\bibfnamefont {G.}~\bibnamefont {Carleo}},\
  }\bibfield  {title} {\bibinfo {title} {An efficient quantum algorithm for the
  time evolution of parameterized circuits},\ }\href
  {https://doi.org/10.22331/q-2021-07-28-512} {\bibfield  {journal} {\bibinfo
  {journal} {{Quantum}}\ }\textbf {\bibinfo {volume} {5}},\ \bibinfo {pages}
  {512} (\bibinfo {year} {2021})}\BibitemShut {NoStop}%
\bibitem [{\citenamefont {Linteau}\ \emph {et~al.}(2024)\citenamefont
  {Linteau}, \citenamefont {Barison}, \citenamefont {Lindner},\ and\
  \citenamefont {Carleo}}]{Linteau2024adaptiveprojectedVQD}%
  \BibitemOpen
  \bibfield  {author} {\bibinfo {author} {\bibfnamefont {D.}~\bibnamefont
  {Linteau}}, \bibinfo {author} {\bibfnamefont {S.}~\bibnamefont {Barison}},
  \bibinfo {author} {\bibfnamefont {N.~H.}\ \bibnamefont {Lindner}},\ and\
  \bibinfo {author} {\bibfnamefont {G.}~\bibnamefont {Carleo}},\ }\bibfield
  {title} {\bibinfo {title} {Adaptive projected variational quantum dynamics},\
  }\href {https://doi.org/10.1103/PhysRevResearch.6.023130} {\bibfield
  {journal} {\bibinfo  {journal} {Phys. Rev. Res.}\ }\textbf {\bibinfo {volume}
  {6}},\ \bibinfo {pages} {023130} (\bibinfo {year} {2024})}\BibitemShut
  {NoStop}%
\bibitem [{\citenamefont {Gibbs}\ \emph {et~al.}(2022)\citenamefont {Gibbs},
  \citenamefont {Gili}, \citenamefont {Holmes}, \citenamefont {Commeau},
  \citenamefont {Arrasmith}, \citenamefont {Cincio}, \citenamefont {Coles},\
  and\ \citenamefont {Sornborger}}]{gibbs2021longtime}%
  \BibitemOpen
  \bibfield  {author} {\bibinfo {author} {\bibfnamefont {J.}~\bibnamefont
  {Gibbs}}, \bibinfo {author} {\bibfnamefont {K.}~\bibnamefont {Gili}},
  \bibinfo {author} {\bibfnamefont {Z.}~\bibnamefont {Holmes}}, \bibinfo
  {author} {\bibfnamefont {B.}~\bibnamefont {Commeau}}, \bibinfo {author}
  {\bibfnamefont {A.}~\bibnamefont {Arrasmith}}, \bibinfo {author}
  {\bibfnamefont {L.}~\bibnamefont {Cincio}}, \bibinfo {author} {\bibfnamefont
  {P.~J.}\ \bibnamefont {Coles}},\ and\ \bibinfo {author} {\bibfnamefont
  {A.}~\bibnamefont {Sornborger}},\ }\bibfield  {title} {\bibinfo {title}
  {Long-time simulations for fixed input states on quantum hardware},\ }\href
  {https://doi.org/10.1038/s41534-022-00625-0} {\bibfield  {journal} {\bibinfo
  {journal} {npj Quantum Information}\ }\textbf {\bibinfo {volume} {8}},\
  \bibinfo {pages} {135} (\bibinfo {year} {2022})}\BibitemShut {NoStop}%
\bibitem [{\citenamefont {Heya}\ \emph {et~al.}(2023)\citenamefont {Heya},
  \citenamefont {Nakanishi}, \citenamefont {Mitarai}, \citenamefont {Yan},
  \citenamefont {Zuo}, \citenamefont {Suzuki}, \citenamefont {Sugiyama},
  \citenamefont {Tamate}, \citenamefont {Tabuchi}, \citenamefont {Fujii},\ and\
  \citenamefont {Nakamura}}]{Heya2023subspaceVQS}%
  \BibitemOpen
  \bibfield  {author} {\bibinfo {author} {\bibfnamefont {K.}~\bibnamefont
  {Heya}}, \bibinfo {author} {\bibfnamefont {K.~M.}\ \bibnamefont {Nakanishi}},
  \bibinfo {author} {\bibfnamefont {K.}~\bibnamefont {Mitarai}}, \bibinfo
  {author} {\bibfnamefont {Z.}~\bibnamefont {Yan}}, \bibinfo {author}
  {\bibfnamefont {K.}~\bibnamefont {Zuo}}, \bibinfo {author} {\bibfnamefont
  {Y.}~\bibnamefont {Suzuki}}, \bibinfo {author} {\bibfnamefont
  {T.}~\bibnamefont {Sugiyama}}, \bibinfo {author} {\bibfnamefont
  {S.}~\bibnamefont {Tamate}}, \bibinfo {author} {\bibfnamefont
  {Y.}~\bibnamefont {Tabuchi}}, \bibinfo {author} {\bibfnamefont
  {K.}~\bibnamefont {Fujii}},\ and\ \bibinfo {author} {\bibfnamefont
  {Y.}~\bibnamefont {Nakamura}},\ }\bibfield  {title} {\bibinfo {title}
  {Subspace variational quantum simulator},\ }\href
  {https://doi.org/10.1103/PhysRevResearch.5.023078} {\bibfield  {journal}
  {\bibinfo  {journal} {Phys. Rev. Res.}\ }\textbf {\bibinfo {volume} {5}},\
  \bibinfo {pages} {023078} (\bibinfo {year} {2023})}\BibitemShut {NoStop}%
\bibitem [{\citenamefont {Cerezo}\ \emph {et~al.}(2021)\citenamefont {Cerezo},
  \citenamefont {Arrasmith}, \citenamefont {Babbush}, \citenamefont {Benjamin},
  \citenamefont {Endo}, \citenamefont {Fujii}, \citenamefont {McClean},
  \citenamefont {Mitarai}, \citenamefont {Yuan}, \citenamefont {Cincio} \emph
  {et~al.}}]{cerezo2021variational}%
  \BibitemOpen
  \bibfield  {author} {\bibinfo {author} {\bibfnamefont {M.}~\bibnamefont
  {Cerezo}}, \bibinfo {author} {\bibfnamefont {A.}~\bibnamefont {Arrasmith}},
  \bibinfo {author} {\bibfnamefont {R.}~\bibnamefont {Babbush}}, \bibinfo
  {author} {\bibfnamefont {S.~C.}\ \bibnamefont {Benjamin}}, \bibinfo {author}
  {\bibfnamefont {S.}~\bibnamefont {Endo}}, \bibinfo {author} {\bibfnamefont
  {K.}~\bibnamefont {Fujii}}, \bibinfo {author} {\bibfnamefont {J.~R.}\
  \bibnamefont {McClean}}, \bibinfo {author} {\bibfnamefont {K.}~\bibnamefont
  {Mitarai}}, \bibinfo {author} {\bibfnamefont {X.}~\bibnamefont {Yuan}},
  \bibinfo {author} {\bibfnamefont {L.}~\bibnamefont {Cincio}}, \emph
  {et~al.},\ }\bibfield  {title} {\bibinfo {title} {Variational quantum
  algorithms},\ }\href {https://doi.org/10.1038/s42254-021-00348-9} {\bibfield
  {journal} {\bibinfo  {journal} {Nat. Rev. Phys.}\ }\textbf {\bibinfo {volume}
  {3}},\ \bibinfo {pages} {625} (\bibinfo {year} {2021})}\BibitemShut {NoStop}%
\bibitem [{\citenamefont {Anastasiou}\ \emph {et~al.}(2022)\citenamefont
  {Anastasiou}, \citenamefont {Chen}, \citenamefont {Mayhall}, \citenamefont
  {Barnes},\ and\ \citenamefont {Economou}}]{Anastasiou2022}%
  \BibitemOpen
  \bibfield  {author} {\bibinfo {author} {\bibfnamefont {P.~G.}\ \bibnamefont
  {Anastasiou}}, \bibinfo {author} {\bibfnamefont {Y.}~\bibnamefont {Chen}},
  \bibinfo {author} {\bibfnamefont {N.~J.}\ \bibnamefont {Mayhall}}, \bibinfo
  {author} {\bibfnamefont {E.}~\bibnamefont {Barnes}},\ and\ \bibinfo {author}
  {\bibfnamefont {S.~E.}\ \bibnamefont {Economou}},\ }\href@noop {} {\bibinfo
  {title} {Tetris-adapt-vqe: An adaptive algorithm that yields shallower,
  denser circuit ans\"atze}} (\bibinfo {year} {2022}),\ \Eprint
  {https://arxiv.org/abs/2209.10562} {arXiv:2209.10562 [quant-ph]} \BibitemShut
  {NoStop}%
\bibitem [{\citenamefont {Hansen}(1990)}]{hansen1990trucated}%
  \BibitemOpen
  \bibfield  {author} {\bibinfo {author} {\bibfnamefont {P.~C.}\ \bibnamefont
  {Hansen}},\ }\bibfield  {title} {\bibinfo {title} {Truncated singular value
  decomposition solutions to discrete ill-posed problems with ill-determined
  numerical rank},\ }\href {https://doi.org/10.1137/0911028} {\bibfield
  {journal} {\bibinfo  {journal} {SIAM Journal on Scientific and Statistical
  Computing}\ }\textbf {\bibinfo {volume} {11}},\ \bibinfo {pages} {503}
  (\bibinfo {year} {1990})}\BibitemShut {NoStop}%
\bibitem [{\citenamefont {Gacon}\ \emph {et~al.}(2023)\citenamefont {Gacon},
  \citenamefont {Zoufal}, \citenamefont {Carleo},\ and\ \citenamefont
  {Woerner}}]{gacon2023ieee}%
  \BibitemOpen
  \bibfield  {author} {\bibinfo {author} {\bibfnamefont {J.}~\bibnamefont
  {Gacon}}, \bibinfo {author} {\bibfnamefont {C.}~\bibnamefont {Zoufal}},
  \bibinfo {author} {\bibfnamefont {G.}~\bibnamefont {Carleo}},\ and\ \bibinfo
  {author} {\bibfnamefont {S.}~\bibnamefont {Woerner}},\ }\bibfield  {title}
  {\bibinfo {title} {Stochastic approximation of variational quantum imaginary
  time evolution},\ }in\ \href {https://doi.org/10.1109/QCE57702.2023.10367741}
  {\emph {\bibinfo {booktitle} {2023 IEEE International Conference on Quantum
  Computing and Engineering (QCE)}}}\ (\bibinfo  {publisher} {IEEE Computer
  Society},\ \bibinfo {address} {Los Alamitos, CA, USA},\ \bibinfo {year}
  {2023})\ pp.\ \bibinfo {pages} {129--139}\BibitemShut {NoStop}%
\bibitem [{\citenamefont {McArdle}\ \emph {et~al.}(2019)\citenamefont
  {McArdle}, \citenamefont {Jones}, \citenamefont {Endo}, \citenamefont {Li},
  \citenamefont {Benjamin},\ and\ \citenamefont {Yuan}}]{VQITE}%
  \BibitemOpen
  \bibfield  {author} {\bibinfo {author} {\bibfnamefont {S.}~\bibnamefont
  {McArdle}}, \bibinfo {author} {\bibfnamefont {T.}~\bibnamefont {Jones}},
  \bibinfo {author} {\bibfnamefont {S.}~\bibnamefont {Endo}}, \bibinfo {author}
  {\bibfnamefont {Y.}~\bibnamefont {Li}}, \bibinfo {author} {\bibfnamefont
  {S.~C.}\ \bibnamefont {Benjamin}},\ and\ \bibinfo {author} {\bibfnamefont
  {X.}~\bibnamefont {Yuan}},\ }\bibfield  {title} {\bibinfo {title}
  {Variational ansatz-based quantum simulation of imaginary time evolution},\
  }\href {https://doi.org/10.1038/s41534-019-0187-2} {\bibfield  {journal}
  {\bibinfo  {journal} {npj Quantum Inf.}\ }\textbf {\bibinfo {volume} {5}},\
  \bibinfo {pages} {75} (\bibinfo {year} {2019})}\BibitemShut {NoStop}%
\bibitem [{\citenamefont {Gomes}\ \emph {et~al.}(2021)\citenamefont {Gomes},
  \citenamefont {Mukherjee}, \citenamefont {Zhang}, \citenamefont {Iadecola},
  \citenamefont {Wang}, \citenamefont {Ho}, \citenamefont {Orth},\ and\
  \citenamefont {Yao}}]{AVQITE}%
  \BibitemOpen
  \bibfield  {author} {\bibinfo {author} {\bibfnamefont {N.}~\bibnamefont
  {Gomes}}, \bibinfo {author} {\bibfnamefont {A.}~\bibnamefont {Mukherjee}},
  \bibinfo {author} {\bibfnamefont {F.}~\bibnamefont {Zhang}}, \bibinfo
  {author} {\bibfnamefont {T.}~\bibnamefont {Iadecola}}, \bibinfo {author}
  {\bibfnamefont {C.-Z.}\ \bibnamefont {Wang}}, \bibinfo {author}
  {\bibfnamefont {K.-M.}\ \bibnamefont {Ho}}, \bibinfo {author} {\bibfnamefont
  {P.~P.}\ \bibnamefont {Orth}},\ and\ \bibinfo {author} {\bibfnamefont
  {Y.-X.}\ \bibnamefont {Yao}},\ }\bibfield  {title} {\bibinfo {title}
  {Adaptive variational quantum imaginary time evolution approach for ground
  state preparation},\ }\href {https://doi.org/10.1002/qute.202100114}
  {\bibfield  {journal} {\bibinfo  {journal} {Adv. Quantum Technol.}\ }\textbf
  {\bibinfo {volume} {4}},\ \bibinfo {pages} {2100114} (\bibinfo {year}
  {2021})}\BibitemShut {NoStop}%
\bibitem [{\citenamefont {Chen}\ \emph {et~al.}(2024)\citenamefont {Chen},
  \citenamefont {Gomes}, \citenamefont {Niu},\ and\ \citenamefont
  {Jong}}]{Chen2024_avqite_open}%
  \BibitemOpen
  \bibfield  {author} {\bibinfo {author} {\bibfnamefont {H.}~\bibnamefont
  {Chen}}, \bibinfo {author} {\bibfnamefont {N.}~\bibnamefont {Gomes}},
  \bibinfo {author} {\bibfnamefont {S.}~\bibnamefont {Niu}},\ and\ \bibinfo
  {author} {\bibfnamefont {W.~A.~d.}\ \bibnamefont {Jong}},\ }\bibfield
  {title} {\bibinfo {title} {Adaptive variational simulation for open quantum
  systems},\ }\href {https://doi.org/10.22331/q-2024-02-13-1252} {\bibfield
  {journal} {\bibinfo  {journal} {{Quantum}}\ }\textbf {\bibinfo {volume}
  {8}},\ \bibinfo {pages} {1252} (\bibinfo {year} {2024})}\BibitemShut
  {NoStop}%
\bibitem [{\citenamefont {Nielsen}\ and\ \citenamefont
  {Chuang}(2011)}]{nielsen2002quantum}%
  \BibitemOpen
  \bibfield  {author} {\bibinfo {author} {\bibfnamefont {M.~A.}\ \bibnamefont
  {Nielsen}}\ and\ \bibinfo {author} {\bibfnamefont {I.~L.}\ \bibnamefont
  {Chuang}},\ }\href {https://doi.org/10.1017/CBO9780511976667} {\emph
  {\bibinfo {title} {Quantum Computation and Quantum Information: 10th
  Anniversary Edition}}},\ \bibinfo {edition} {10th}\ ed.\ (\bibinfo
  {publisher} {Cambridge University Press},\ \bibinfo {address} {New York,
  USA},\ \bibinfo {year} {2011})\BibitemShut {NoStop}%
\bibitem [{\citenamefont {Iserles}(2009)}]{iserles2009ode}%
  \BibitemOpen
  \bibfield  {author} {\bibinfo {author} {\bibfnamefont {A.}~\bibnamefont
  {Iserles}},\ }\href {https://doi.org/10.1119/1.18632} {\emph {\bibinfo
  {title} {A first course in the numerical analysis of differential
  equations}}},\ \bibinfo {number} {44}\ (\bibinfo  {publisher} {Cambridge
  university press},\ \bibinfo {address} {New York, USA},\ \bibinfo {year}
  {2009})\BibitemShut {NoStop}%
\bibitem [{\citenamefont {Trotter}(1959)}]{trotter}%
  \BibitemOpen
  \bibfield  {author} {\bibinfo {author} {\bibfnamefont {H.~F.}\ \bibnamefont
  {Trotter}},\ }\bibfield  {title} {\bibinfo {title} {On the product of
  semi-groups of operators},\ }\href
  {https://doi.org/10.1090/S0002-9939-1959-0108732-6} {\bibfield  {journal}
  {\bibinfo  {journal} {Proc. Am. Math. Soc.}\ }\textbf {\bibinfo {volume}
  {10}},\ \bibinfo {pages} {545} (\bibinfo {year} {1959})}\BibitemShut
  {NoStop}%
\bibitem [{\citenamefont {Wecker}\ \emph
  {et~al.}(2015{\natexlab{b}})\citenamefont {Wecker}, \citenamefont
  {Hastings},\ and\ \citenamefont {Troyer}}]{wecker2015_trotterizedsp}%
  \BibitemOpen
  \bibfield  {author} {\bibinfo {author} {\bibfnamefont {D.}~\bibnamefont
  {Wecker}}, \bibinfo {author} {\bibfnamefont {M.~B.}\ \bibnamefont
  {Hastings}},\ and\ \bibinfo {author} {\bibfnamefont {M.}~\bibnamefont
  {Troyer}},\ }\bibfield  {title} {\bibinfo {title} {Progress towards practical
  quantum variational algorithms},\ }\href
  {https://doi.org/10.1103/PhysRevA.92.042303} {\bibfield  {journal} {\bibinfo
  {journal} {Phys. Rev. A}\ }\textbf {\bibinfo {volume} {92}},\ \bibinfo
  {pages} {042303} (\bibinfo {year} {2015}{\natexlab{b}})}\BibitemShut
  {NoStop}%
\bibitem [{\citenamefont {Barison}\ \emph {et~al.}(2022)\citenamefont
  {Barison}, \citenamefont {Vicentini}, \citenamefont {Cirac},\ and\
  \citenamefont {Carleo}}]{barison2022variational}%
  \BibitemOpen
  \bibfield  {author} {\bibinfo {author} {\bibfnamefont {S.}~\bibnamefont
  {Barison}}, \bibinfo {author} {\bibfnamefont {F.}~\bibnamefont {Vicentini}},
  \bibinfo {author} {\bibfnamefont {I.}~\bibnamefont {Cirac}},\ and\ \bibinfo
  {author} {\bibfnamefont {G.}~\bibnamefont {Carleo}},\ }\bibfield  {title}
  {\bibinfo {title} {Variational dynamics as a ground-state problem on a
  quantum computer},\ }\href {https://doi.org/10.1103/PhysRevResearch.4.043161}
  {\bibfield  {journal} {\bibinfo  {journal} {Phys. Rev. Res.}\ }\textbf
  {\bibinfo {volume} {4}},\ \bibinfo {pages} {043161} (\bibinfo {year}
  {2022})}\BibitemShut {NoStop}%
\bibitem [{\citenamefont {Mukherjee}\ \emph {et~al.}(2023)\citenamefont
  {Mukherjee}, \citenamefont {Berthusen}, \citenamefont {Getelina},
  \citenamefont {Orth},\ and\ \citenamefont {Yao}}]{mukherjee2023comparative}%
  \BibitemOpen
  \bibfield  {author} {\bibinfo {author} {\bibfnamefont {A.}~\bibnamefont
  {Mukherjee}}, \bibinfo {author} {\bibfnamefont {N.~F.}\ \bibnamefont
  {Berthusen}}, \bibinfo {author} {\bibfnamefont {J.~C.}\ \bibnamefont
  {Getelina}}, \bibinfo {author} {\bibfnamefont {P.~P.}\ \bibnamefont {Orth}},\
  and\ \bibinfo {author} {\bibfnamefont {Y.-X.}\ \bibnamefont {Yao}},\
  }\bibfield  {title} {\bibinfo {title} {Comparative study of adaptive
  variational quantum eigensolvers for multi-orbital impurity models},\ }\href
  {https://doi.org/10.1038/s42005-022-01089-6} {\bibfield  {journal} {\bibinfo
  {journal} {Commun. Phys.}\ }\textbf {\bibinfo {volume} {6}},\ \bibinfo
  {pages} {4} (\bibinfo {year} {2023})}\BibitemShut {NoStop}%
\bibitem [{\citenamefont {Abraham}\ \emph {et~al.}(2019)\citenamefont
  {Abraham}, \citenamefont {Akhalwaya}, \citenamefont {Aleksandrowicz},
  \citenamefont {Alexander}, \citenamefont {Alexandrowics}, \citenamefont
  {Arbel}, \citenamefont {Asfaw}, \citenamefont {Azaustre}, \citenamefont
  {AzizNgoueya}, \citenamefont {Barkoutsos}, \citenamefont {Barron},
  \citenamefont {Bello}, \citenamefont {Ben-Haim}, \citenamefont {Bevenius}
  \emph {et~al.}}]{Qiskit}%
  \BibitemOpen
  \bibfield  {author} {\bibinfo {author} {\bibfnamefont {H.}~\bibnamefont
  {Abraham}}, \bibinfo {author} {\bibfnamefont {I.~Y.}\ \bibnamefont
  {Akhalwaya}}, \bibinfo {author} {\bibfnamefont {G.}~\bibnamefont
  {Aleksandrowicz}}, \bibinfo {author} {\bibfnamefont {T.}~\bibnamefont
  {Alexander}}, \bibinfo {author} {\bibfnamefont {G.}~\bibnamefont
  {Alexandrowics}}, \bibinfo {author} {\bibfnamefont {E.}~\bibnamefont
  {Arbel}}, \bibinfo {author} {\bibfnamefont {A.}~\bibnamefont {Asfaw}},
  \bibinfo {author} {\bibfnamefont {C.}~\bibnamefont {Azaustre}}, \bibinfo
  {author} {\bibnamefont {AzizNgoueya}}, \bibinfo {author} {\bibfnamefont
  {P.}~\bibnamefont {Barkoutsos}}, \bibinfo {author} {\bibfnamefont
  {G.}~\bibnamefont {Barron}}, \bibinfo {author} {\bibfnamefont
  {L.}~\bibnamefont {Bello}}, \bibinfo {author} {\bibfnamefont
  {Y.}~\bibnamefont {Ben-Haim}}, \bibinfo {author} {\bibfnamefont
  {D.}~\bibnamefont {Bevenius}}, \emph {et~al.},\ }\href@noop {} {\bibinfo
  {title} {Qiskit: An open-source framework for quantum computing}} (\bibinfo
  {year} {2019})\BibitemShut {NoStop}%
\bibitem [{pvq()}]{pvqd_github}%
  \BibitemOpen
  \bibfield  {title} {\bibinfo {title} {Adaptive projected variational quantum
  dynamics (adaptive pvqd)},\ }\href@noop {} {\bibinfo  {journal}
  {https://github.com/dalin27/adaptive-pvqd}\ }\BibitemShut {NoStop}%
\bibitem [{\citenamefont {Meyer}(2021)}]{meyer2021fisher}%
  \BibitemOpen
\bibfield  {journal} {  }\bibfield  {author} {\bibinfo {author} {\bibfnamefont
  {J.~J.}\ \bibnamefont {Meyer}},\ }\bibfield  {title} {\bibinfo {title}
  {Fisher information in noisy intermediate-scale quantum applications},\
  }\href {https://doi.org/10.22331/q-2021-09-09-539} {\bibfield  {journal}
  {\bibinfo  {journal} {Quantum}\ }\textbf {\bibinfo {volume} {5}},\ \bibinfo
  {pages} {539} (\bibinfo {year} {2021})}\BibitemShut {NoStop}%
\bibitem [{\citenamefont {Kolotouros}\ \emph {et~al.}(2024)\citenamefont
  {Kolotouros}, \citenamefont {Joseph},\ and\ \citenamefont
  {Narayanan}}]{kolotouros2024acceleratingquantumimaginarytimeevolution}%
  \BibitemOpen
  \bibfield  {author} {\bibinfo {author} {\bibfnamefont {I.}~\bibnamefont
  {Kolotouros}}, \bibinfo {author} {\bibfnamefont {D.}~\bibnamefont {Joseph}},\
  and\ \bibinfo {author} {\bibfnamefont {A.~K.}\ \bibnamefont {Narayanan}},\
  }\href {https://arxiv.org/abs/2407.03123} {\bibinfo {title} {Accelerating
  quantum imaginary-time evolution with random measurements}} (\bibinfo {year}
  {2024}),\ \Eprint {https://arxiv.org/abs/2407.03123} {arXiv:2407.03123
  [quant-ph]} \BibitemShut {NoStop}%
\bibitem [{\citenamefont {Liu}\ \emph {et~al.}(2019)\citenamefont {Liu},
  \citenamefont {Yuan}, \citenamefont {Lu},\ and\ \citenamefont
  {Wang}}]{Liu2020QuantumFI}%
  \BibitemOpen
  \bibfield  {author} {\bibinfo {author} {\bibfnamefont {J.}~\bibnamefont
  {Liu}}, \bibinfo {author} {\bibfnamefont {H.}~\bibnamefont {Yuan}}, \bibinfo
  {author} {\bibfnamefont {X.-M.}\ \bibnamefont {Lu}},\ and\ \bibinfo {author}
  {\bibfnamefont {X.}~\bibnamefont {Wang}},\ }\bibfield  {title} {\bibinfo
  {title} {Quantum fisher information matrix and multiparameter estimation},\
  }\href {https://doi.org/10.1088/1751-8121/ab5d4d} {\bibfield  {journal}
  {\bibinfo  {journal} {Journal of Physics A: Mathematical and Theoretical}\
  }\textbf {\bibinfo {volume} {53}},\ \bibinfo {pages} {023001} (\bibinfo
  {year} {2019})}\BibitemShut {NoStop}%
\bibitem [{\citenamefont {Gacon}\ \emph {et~al.}(2021)\citenamefont {Gacon},
  \citenamefont {Zoufal}, \citenamefont {Carleo},\ and\ \citenamefont
  {Woerner}}]{Gacon2021simultaneousPS}%
  \BibitemOpen
  \bibfield  {author} {\bibinfo {author} {\bibfnamefont {J.}~\bibnamefont
  {Gacon}}, \bibinfo {author} {\bibfnamefont {C.}~\bibnamefont {Zoufal}},
  \bibinfo {author} {\bibfnamefont {G.}~\bibnamefont {Carleo}},\ and\ \bibinfo
  {author} {\bibfnamefont {S.}~\bibnamefont {Woerner}},\ }\bibfield  {title}
  {\bibinfo {title} {Simultaneous {P}erturbation {S}tochastic {A}pproximation
  of the {Q}uantum {F}isher {I}nformation},\ }\href
  {https://doi.org/10.22331/q-2021-10-20-567} {\bibfield  {journal} {\bibinfo
  {journal} {{Quantum}}\ }\textbf {\bibinfo {volume} {5}},\ \bibinfo {pages}
  {567} (\bibinfo {year} {2021})}\BibitemShut {NoStop}%
\bibitem [{\citenamefont {Ayral}\ \emph {et~al.}(2023)\citenamefont {Ayral},
  \citenamefont {Louvet}, \citenamefont {Zhou}, \citenamefont {Lambert},
  \citenamefont {Stoudenmire},\ and\ \citenamefont {Waintal}}]{Ayral2023}%
  \BibitemOpen
  \bibfield  {author} {\bibinfo {author} {\bibfnamefont {T.}~\bibnamefont
  {Ayral}}, \bibinfo {author} {\bibfnamefont {T.}~\bibnamefont {Louvet}},
  \bibinfo {author} {\bibfnamefont {Y.}~\bibnamefont {Zhou}}, \bibinfo {author}
  {\bibfnamefont {C.}~\bibnamefont {Lambert}}, \bibinfo {author} {\bibfnamefont
  {E.~M.}\ \bibnamefont {Stoudenmire}},\ and\ \bibinfo {author} {\bibfnamefont
  {X.}~\bibnamefont {Waintal}},\ }\bibfield  {title} {\bibinfo {title}
  {Density-matrix renormalization group algorithm for simulating quantum
  circuits with a finite fidelity},\ }\bibfield  {journal} {\bibinfo  {journal}
  {PRX Quantum}\ }\textbf {\bibinfo {volume} {4}},\ \href
  {https://doi.org/10.1103/PRXQuantum.4.020304} {10.1103/PRXQuantum.4.020304}
  (\bibinfo {year} {2023})\BibitemShut {NoStop}%
\bibitem [{\citenamefont {Tindall}\ \emph {et~al.}(2024)\citenamefont
  {Tindall}, \citenamefont {Fishman}, \citenamefont {Stoudenmire},\ and\
  \citenamefont {Sels}}]{Tindall2024EfficientTN}%
  \BibitemOpen
  \bibfield  {author} {\bibinfo {author} {\bibfnamefont {J.}~\bibnamefont
  {Tindall}}, \bibinfo {author} {\bibfnamefont {M.}~\bibnamefont {Fishman}},
  \bibinfo {author} {\bibfnamefont {E.~M.}\ \bibnamefont {Stoudenmire}},\ and\
  \bibinfo {author} {\bibfnamefont {D.}~\bibnamefont {Sels}},\ }\bibfield
  {title} {\bibinfo {title} {Efficient tensor network simulation of ibm’s
  eagle kicked ising experiment},\ }\bibfield  {journal} {\bibinfo  {journal}
  {PRX Quantum}\ }\textbf {\bibinfo {volume} {5}},\ \href
  {https://doi.org/10.1103/prxquantum.5.010308} {10.1103/prxquantum.5.010308}
  (\bibinfo {year} {2024})\BibitemShut {NoStop}%
\bibitem [{\citenamefont {Herrmann}\ \emph {et~al.}(2023)\citenamefont
  {Herrmann}, \citenamefont {Arya}, \citenamefont {Doherty}, \citenamefont
  {Mingare}, \citenamefont {Pillay}, \citenamefont {Preis},\ and\ \citenamefont
  {Prestel}}]{herrmann2023quantumud}%
  \BibitemOpen
  \bibfield  {author} {\bibinfo {author} {\bibfnamefont {N.}~\bibnamefont
  {Herrmann}}, \bibinfo {author} {\bibfnamefont {D.}~\bibnamefont {Arya}},
  \bibinfo {author} {\bibfnamefont {M.~W.}\ \bibnamefont {Doherty}}, \bibinfo
  {author} {\bibfnamefont {A.}~\bibnamefont {Mingare}}, \bibinfo {author}
  {\bibfnamefont {J.~C.}\ \bibnamefont {Pillay}}, \bibinfo {author}
  {\bibfnamefont {F.}~\bibnamefont {Preis}},\ and\ \bibinfo {author}
  {\bibfnamefont {S.}~\bibnamefont {Prestel}},\ }\bibfield  {title} {\bibinfo
  {title} {Quantum utility – definition and assessment of a practical quantum
  advantage},\ }in\ \href {https://doi.org/10.1109/qsw59989.2023.00028} {\emph
  {\bibinfo {booktitle} {2023 IEEE International Conference on Quantum Software
  (QSW)}}}\ (\bibinfo  {publisher} {IEEE},\ \bibinfo {year} {2023})\BibitemShut
  {NoStop}%
\bibitem [{\citenamefont {Kim}\ \emph {et~al.}(2023)\citenamefont {Kim},
  \citenamefont {Eddins}, \citenamefont {Anand}, \citenamefont {Wei},
  \citenamefont {{van den Berg}}, \citenamefont {Rosenblatt}, \citenamefont
  {Nayfeh}, \citenamefont {Wu}, \citenamefont {Zaletel}, \citenamefont
  {Temme},\ and\ \citenamefont {Kandala}}]{kimEvidenceUtilityQuantum2023}%
  \BibitemOpen
  \bibfield  {author} {\bibinfo {author} {\bibfnamefont {Y.}~\bibnamefont
  {Kim}}, \bibinfo {author} {\bibfnamefont {A.}~\bibnamefont {Eddins}},
  \bibinfo {author} {\bibfnamefont {S.}~\bibnamefont {Anand}}, \bibinfo
  {author} {\bibfnamefont {K.~X.}\ \bibnamefont {Wei}}, \bibinfo {author}
  {\bibfnamefont {E.}~\bibnamefont {{van den Berg}}}, \bibinfo {author}
  {\bibfnamefont {S.}~\bibnamefont {Rosenblatt}}, \bibinfo {author}
  {\bibfnamefont {H.}~\bibnamefont {Nayfeh}}, \bibinfo {author} {\bibfnamefont
  {Y.}~\bibnamefont {Wu}}, \bibinfo {author} {\bibfnamefont {M.}~\bibnamefont
  {Zaletel}}, \bibinfo {author} {\bibfnamefont {K.}~\bibnamefont {Temme}},\
  and\ \bibinfo {author} {\bibfnamefont {A.}~\bibnamefont {Kandala}},\
  }\bibfield  {title} {\bibinfo {title} {Evidence for the utility of quantum
  computing before fault tolerance},\ }\href
  {https://doi.org/10.1038/s41586-023-06096-3} {\bibfield  {journal} {\bibinfo
  {journal} {Nature}\ }\textbf {\bibinfo {volume} {618}},\ \bibinfo {pages}
  {500} (\bibinfo {year} {2023})}\BibitemShut {NoStop}%
\bibitem [{\citenamefont {Endo}\ \emph
  {et~al.}(2020{\natexlab{b}})\citenamefont {Endo}, \citenamefont {Kurata},\
  and\ \citenamefont {Nakagawa}}]{endo2020calculation}%
  \BibitemOpen
  \bibfield  {author} {\bibinfo {author} {\bibfnamefont {S.}~\bibnamefont
  {Endo}}, \bibinfo {author} {\bibfnamefont {I.}~\bibnamefont {Kurata}},\ and\
  \bibinfo {author} {\bibfnamefont {Y.~O.}\ \bibnamefont {Nakagawa}},\
  }\bibfield  {title} {\bibinfo {title} {Calculation of the green's function on
  near-term quantum computers},\ }\href
  {https://doi.org/10.1103/PhysRevResearch.2.033281} {\bibfield  {journal}
  {\bibinfo  {journal} {Phys. Rev. Res.}\ }\textbf {\bibinfo {volume} {2}},\
  \bibinfo {pages} {033281} (\bibinfo {year} {2020}{\natexlab{b}})}\BibitemShut
  {NoStop}%
\bibitem [{\citenamefont {Hanke}\ and\ \citenamefont
  {Hansen}(1993)}]{hanke1989regularization}%
  \BibitemOpen
  \bibfield  {author} {\bibinfo {author} {\bibfnamefont {M.}~\bibnamefont
  {Hanke}}\ and\ \bibinfo {author} {\bibfnamefont {P.~C.}\ \bibnamefont
  {Hansen}},\ }\bibfield  {title} {\bibinfo {title} {Regularization methods for
  large-scale problems},\ }\href@noop {} {\bibfield  {journal} {\bibinfo
  {journal} {Surveys on Mathematics for Industry}\ }\textbf {\bibinfo {volume}
  {3}},\ \bibinfo {pages} {253} (\bibinfo {year} {1993})}\BibitemShut {NoStop}%
\end{thebibliography}
%

\end{document}